\documentclass{aastex}

\usepackage[maxfloats=28]{morefloats}
\usepackage{amsmath}
\usepackage{amstext}
\usepackage{graphicx}
\usepackage[caption=false]{subfig}
\usepackage[usenames, dvipsnames]{color}
\usepackage{soul}

\begin{document}

%Title
\title{The effect of rotation on oscillatory double-diffusive convection (semiconvection)}
\author{Ryan Moll and Pascale Garaud}
\affil{Department of Applied Mathematics and Statistics, Baskin School of Engineering, University of California Santa Cruz, CA 95064, USA}
\email{rmoll@soe.ucsc.edu}
%Abstract
\begin{abstract}
Oscillatory double-diffusive convection (ODDC, more traditionally called semiconvection) is a form of linear double-diffusive instability that occurs in fluids that are unstably stratified in temperature (Schwarzschild unstable), but stably stratified in chemical composition (Ledoux stable). This scenario is thought to be quite common in the interiors of stars and giant planets, and understanding the transport of heat and chemical species by ODDC is of great importance to stellar and planetary evolution models. Fluids unstable to ODDC have a tendency to form convective thermo-compositional layers which significantly enhance the fluxes of temperature and chemical composition compared with microscopic diffusion. Although a number of recent studies have focused on studying properties of both layered and non-layered ODDC, few have addressed how additional physical processes such as global rotation affect its dynamics. In this work we study first how rotation affects the linear stability properties of rotating ODDC. Using direct numerical simulations we then analyze the effect of rotation on properties of layered and non-layered ODDC, and study how the angle of the rotation axis with respect to the direction of gravity affects layering. We find that rotating systems can be broadly grouped into two categories, based on the strength of rotation. Qualitative behavior in the more weakly rotating group is similar to non-rotating ODDC, but strongly rotating systems become dominated by vortices that are invariant in the direction of the rotation vector and strongly influence transport. We find that whenever layers form, rotation always acts to reduce thermal and compositional transport.
\end{abstract}

%Section: Introduction
\section{Introduction} \label{sec:Intro}
In the gaseous interiors of stars and giant planets, regions that are unstably stratified in temperature (Schwarzschild unstable) but stably stratified in chemical composition (Ledoux stable) are likely to be common. Fluids stratified in this way are, by definition, stable to the sort of overturning motion that occurs in standard convection. However, \citet{walin1964} and \citet{kato1966} showed that, given the right conditions, infinitesimal perturbations can trigger an instability which takes the form of over-stable gravity waves. This instability, often known as semiconvection but more accurately described as oscillatory double-diffusive convection (ODDC) after \citet{spiegel1969}, can lead to significant augmentation of the turbulent transport of temperature and chemical species through a fluid, and is therefore an important process to consider in evolution models of stars and giant planets.

%\begin{equation} \label{eq:nablaR}
%R_0^{-1}=\frac{\frac{\phi}{\delta}\nabla_{\mu}}{\nabla-\nabla_{\rm ad}} \equiv \frac{\frac{\phi}{\delta} \left( \frac{d \ln \mu}{d \ln p} \right)}{\left( \frac{d \ln T}{d \ln p} \right) - \left( \frac{d \ln T}{d \ln p} \right)_{\rm ad}}
%\end{equation}
%where $T$, $p$ and $\mu$ are background profiles of temperature, pressure and chemical composition, respectively, and where $\delta=\left.\frac{\partial \ln \rho}{\partial \ln T}\right|_{p,\mu}$ and $\phi=\left.\frac{\partial \ln \rho}{\partial \ln \mu}\right|_{p,T}$ are thermodynamic derivatives of the equation of state.

Double-diffusive fluids with the kind of stratification described here were first discussed in the geophysical scientific community in the context of volcanic lakes \citep{Newman1976} and the polar ocean \citep{Timmermans2003,toole2006}. There, they became well-known for their propensity to form density staircases consisting of convectively mixed layers separated by stably stratified interfaces. As a result, layered convection is usually studied in experiments where a layered configuration is imposed as an initial condition, rather than following naturally from the growth and non-linear saturation of ODDC. Thermo-compositional layering was first studied in laboratory experiments involving salt water \citep{Turner1965,lindenshirtcliffe1978}, or aqueous sugar/salt solutions \citep{shirtcliffe1973}, that were initialized with layers. The results from these studies were then used to inform studies of double-diffusive fluids in stars \citep{langer1985,merryfield1995} and giant planets \citep{stevenson1982,leconte2012,nettelmann2015}.

However, recent studies have taken a different approach to characterize the dynamics of double-diffusive layering. Advances in high-performance computing have made it feasible to study ODDC using 3D numerical simulations. \citet{rosenblum2011} discovered that layers may form spontaneously in a linearly unstable system, and proposed a mechanism to explain how layer formation occurs. This mechanism, known as the $\gamma-$instability, was originally put forward by \citet{radko2003mlf} to explain layer formation in fingering convection but was found to apply to ODDC as well. The simulations of \citet{rosenblum2011} also demonstrated the existence of a non-layered phase of ODDC which had been neglected by nearly all previous studies except that of \citet{langer1985}, who proposed a model for mixing of chemical species by semiconvection that ignores layering entirely \citep[see reviews by][]{merryfield1995,Moll2016}. Next, \citet{Mirouh2012} identified the parameter regimes in which layers do and do not form by the $\gamma-$instability in ODDC. \citet{Wood2013} then studied the thermal and compositional fluxes through layered ODDC, and \citet{Moll2016} studied the transport characteristics through non-layered ODDC.

In each of these studies, a fairly simple model was used in which the only body force considered was gravity. It is natural to wonder how additional physical mechanisms may affect the long term dynamics of ODDC. Global rotation is one such mechanism that is particularly relevant to the gas giant planets in our own solar system due to their rapid rotation periods ($\sim 9.9$ hours for Jupiter and $\sim 10.7$ hours for Saturn). It is also potentially important to rapidly rotating extra-solar giant planets, and massive stars. There have been some recent studies of rotating layered convection in double-diffusive fluids, but only for the geophysical parameter regime \citep{CarpTimm2014} in conditions that are not unstable to ODDC (or to the $\gamma-$instability). 

In this work we study the effect of global rotation on the linear stability properties and long-term dynamics associated with ODDC. In Section \ref{sec:mathMod} we introduce our mathematical model and in Section \ref{sec:LinStab} we study how rotation affects its linear stability properties. We analyze the impact of Coriolis forces on the formation of thermo-compositional layers in Section \ref{sec:thetaZero} by studying a suite of simulations with parameter values selected to induce layer formation in non-rotating ODDC. In Section \ref{sec:diffParams} we show results from two other sets of simulations at different values of the diffusivities and of the background stratification and study how rotation affects the dynamics of the non-layered phase of ODDC. In Section \ref{sec:incSims} we study the effect of colatitude on layer formation. Finally, in Section \ref{sec:conclusion} we discuss our results and present preliminary conclusions.

%Section: Mathematical Model
\section{Mathematical Model} \label{sec:mathMod}

The basic model assumptions for rotating ODDC are similar to those made in previous studies of the non-rotating systems \citep{rosenblum2011,Mirouh2012,Wood2013,Moll2016}. As in previous work, we consider a domain that is significantly smaller than a density scale height, and where flow speeds are significantly smaller than the sound speed of the medium. This allows us to use the Boussinesq approximation \citep{spiegelveronis1960} and to ignore the effects of curvature. We consider a 3D Cartesian domain centered at radius $r=r_0$, and oriented in such a way that the $z$-axis is in the radial direction, the $x$-axis is aligned with the azimuthal direction, and the $y$-axis is aligned with the meridional direction. We also assume constant background gradients of temperature, $T_{0z}$, and chemical composition, $\mu_{0z}$, over the vertical extent of the box, which are defined as follows:
\begin{eqnarray}
T_{0z} = \frac{\partial T}{\partial r} = \frac{T}{p} \frac{\partial p}{\partial r} \nabla \, , \nonumber \\
\mu_{0z} =\frac{\partial \mu}{\partial r}   = \frac{\mu}{p} \frac{\partial p}{\partial r} \nabla_{\mu} \, ,
\end{eqnarray}
where all the quantities are taken at $r =r_0$. Here, $p$ denotes pressure, $T$ is temperature, $\mu$ is the mean molecular weight, and $\nabla$ and $\nabla_{\mu}$ have their usual astrophysical definitions:
\begin{equation}
\nabla = \frac{d \ln T}{d \ln p} \mbox{   ,   } \nabla_\mu = \frac{d \ln \mu}{d \ln p} \, \mbox{ at } r = r_0 \, .
\end{equation} 

We use a linearized equation of state in which perturbations to the background density profile, $\tilde{\rho}$, are given by
\begin{equation}
\frac{\tilde{\rho}}{\rho_0} = -\alpha \tilde{T} + \beta \tilde{\mu} \, ,
\end{equation}
where $\tilde{T}$, and $\tilde{\mu}$ are perturbations to the background profiles of temperature and chemical composition, respectively, and $\rho_0$ is the mean density of the domain. The coefficient of thermal expansion, $\alpha$, and of compositional contraction, $\beta$, are defined as
\begin{eqnarray}
\alpha &= &-\frac{1}{\rho_0} \left.\frac{\partial \rho}{\partial T}\right|_{p,\mu} \, , \nonumber \\
\beta &= &\frac{1}{\rho_0} \left.\frac{\partial \rho}{\partial \mu}\right|_{p,T} \, .
\end{eqnarray}

We take the effect of rotation into account by assuming that the rotation vector is given by:
\begin{equation} \label{eq:RotAxis}
\mathbf{\Omega} = \left| \mathbf{\Omega} \right| \left( 0,\sin{\theta},\cos{\theta} \right) \, ,
\end{equation}
where $\theta$ is the angle between the rotation axis and the $z-$axis. With this assumed rotation vector, a domain placed at the poles has a rotation axis aligned with the $z$-direction ($\theta=0$), while at the equator the rotation axis is in the $y$-direction ($\theta=\frac{\pi}{2}$). Due the small sizes of the domains considered (compared to a stellar or planetary radius) we use an $f$-plane approximation where rotation is assumed to be constant throughout the domain.

In what follows we use new units for length, $[l]$, time, $[t]$, temperature, $[T]$, and chemical composition, $[\mu]$ as,
\begin{eqnarray} \label{eq:nondim}
[l] &=& d = \left( \frac{\kappa_T \nu}{\alpha g \left| T_{0z} - T_{0z}^{\rm ad} \right|} \right)^{\frac{1}{4}} = \left( \frac{\kappa_T \nu}{\alpha g \frac{T}{p} \left| \frac{\partial p}{\partial r} \right| \left| \nabla - \nabla_{\rm ad} \right|} \right)^{\frac{1}{4}} \, , \nonumber \\
{[t]} &= &\frac{d^2}{\kappa_T} \, , \nonumber \\
{[T]} &= &d \left| T_{0z} - T_{0z}^{\rm ad} \right| \, , \nonumber \\
{[\mu]} &= &\frac{\alpha}{\beta} d \left| T_{0z} - T_{0z}^{\rm ad} \right| \, ,
\end{eqnarray}
where $g$ is the local gravitational acceleration, $\nu$ is the local viscosity, $\kappa_T$ is the local thermal diffusivity, and where $T_{0z}^{\rm ad}$ is the adiabatic temperature gradient defined as
\begin{equation}
T_{0z}^{\rm ad} = \frac{T}{p} \frac{dp}{dr} \nabla_{\rm ad} \mbox{  at  } r = r_0\, .
\end{equation}
The non-dimensional governing equations for rotating ODDC are then given by
\begin{eqnarray} \label{eq:GovEq}
\nabla \cdot \mathbf{u} &= &0  \, , \nonumber \\
\frac{\partial \mathbf{u}}{\partial t} + \mathbf{u} \cdot \nabla \mathbf{u} &= &-{\rm Pr}\nabla \tilde{p} + {\rm Pr}\left( \tilde{T} - \tilde{\mu} \right)\mathbf{\hat{e}}_z + {\rm Pr}\nabla^2 \mathbf{u} - \sqrt{\rm Ta^*} \left( \frac{\mathbf{\Omega}}{\left| \mathbf{\Omega} \right|} \times \mathbf{u} \right) \, , \nonumber \\
\frac{\partial \tilde{T}}{\partial t} + \mathbf{u} \cdot \nabla \tilde{T} - w &= &\nabla^2 \tilde{T} \, , \nonumber \\
\frac{\partial \tilde{\mu}}{\partial t} + \mathbf{u} \cdot \nabla \tilde{\mu} - R_0^{-1} w &= &\tau \nabla^2 \tilde{\mu} \, ,
\end{eqnarray}
where $\mathbf{u} = (u,v,w)$ is the velocity field. This introduces the usual non-dimensional diffusion parameters ${\rm Pr}$ (the Prandtl number) and $\tau$ (the diffusivity ratio) as \begin{equation}
{\rm Pr}=\frac{\nu}{\kappa_T} \: , \: \tau=\frac{\kappa_{\mu}}{\kappa_T} \, ,
\end{equation}
where $\kappa_{\mu}$ is the compositional diffusivity, and the inverse density ratio, $R_0^{-1}$,  as
\begin{equation}
R_0^{-1} = \frac{\beta \left| \mu_{0z} \right|}{\alpha \left| T_{0z} - T_{0z}^{\rm ad} \right|} \, .
\end{equation}

In a non-rotating model ${\rm Pr}$, $\tau$ and $R_0^{-1}$ are sufficient to fully describe the system. In a rotating model though, we must introduce a fourth non-dimensional parameter that controls the strength of rotation, 
\begin{equation}
{\rm Ta^*} = \frac{4 \left| \mathbf{\Omega} \right|^2 d^4}{\kappa_T^2} \, ,
\end{equation}
which is related to the commonly defined Taylor number in studies of rotating Rayleigh-B\'enard convection as:
\begin{equation} \label{eq:TaNum}
{\rm Ta} = \frac{4 \left| \mathbf{\Omega} \right|^2 L_z^4}{\nu^2} = {\rm Pr}^{-2} \left( \frac{L_z}{d} \right)^4 {\rm Ta}^* \, .
\end{equation}

\begin{figure}[h] %fig 1
\includegraphics[width=\linewidth]{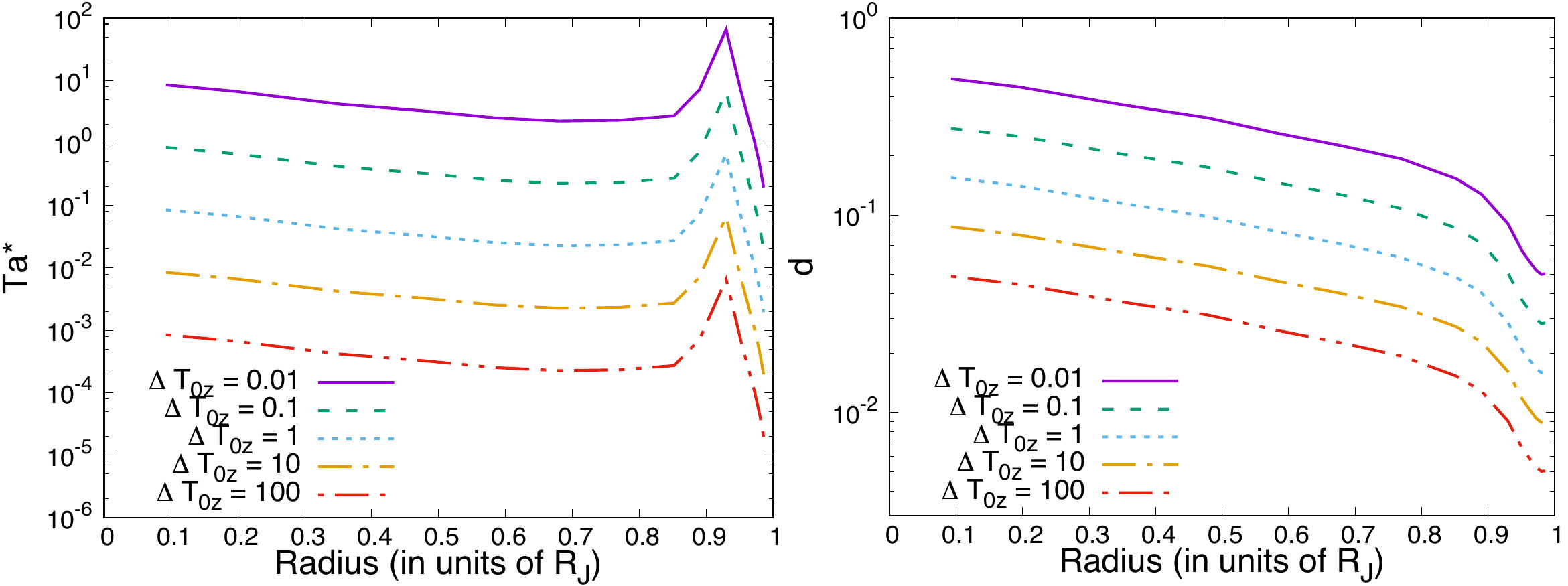}
\caption{Values of  ${\rm Ta^*}$ (left) and $d$ in units of meters (right) estimated for the interior Jupiter using data from \citet{french2012}. Estimates are made for various values of $\Delta T_{0z}$ between $10^{-2}$ and $10^2$.}
 \label{fig:RealTaylor}
\end{figure}

Values of ${\rm Ta^*}$ and $d$ in a stellar or planetary interior are difficult to estimate due to uncertainty in the superadiabaticity of double-diffusive regions. However we can make reasonable estimates for their upper and lower bounds. As in \citet{nettelmann2015} we define the superadiabaticity, $\Delta T_{0z}$, as
\begin{equation}
\Delta T_{0z} = \frac{\nabla - \nabla_{\rm ad}}{\nabla_{\rm ad}} = \frac{ T_{0z} - T_{0z}^{\rm ad} }{ T_{0z}^{\rm ad} } \, .
\end{equation}
In their study values of $\Delta T_{0z}$ were typically between $10^{-2}$ and $10^2$ (see their Figure 2). This range of values, combined with data from \citet{french2012}, allows us to calculate $d$ and ${\rm Ta^*}$ as a function of depth for the interior of Jupiter. From Figure \ref{fig:RealTaylor} we see that the lowest estimates for ${\rm Ta^*}$ are on the order of $10^{-3}$ (for large $\Delta T_{0z}$) and the upper bound is between $1$ and $10$ (for small $\Delta T_{0z}$). As we will show later, this range includes values of ${\rm Ta^*}$ which indicate significant rotational effects on the dynamics of ODDC. Larger values of $\Delta T_{0z}$ are expected in the case of layered ODDC, where $T_{0z}$ is close to $T_{0z}^{\rm ad}$, while smaller values are expected in the case of non-layered ODDC, where $T_{0z}$ is closer to the radiative temperature gradient. 

The conditions for ODDC to occur in a non-rotating fluid are defined by ${\rm Pr}$, $\tau$, and, most importantly, $R_0^{-1}$ \citep{baines1969}. For a system to be unstable to infinitesimal perturbations, $R_0^{-1}$ must be within the following range:
\begin{equation} \label{eq:LinStabCrit}
1 < R_0^{-1} < R_c^{-1} \equiv \frac{{\rm Pr}+1}{{\rm Pr}+\tau} \, .
\end{equation}
If $R_0^{-1} < 1$, the system is unstable to standard convection, and if $R_0^{-1} > R_c^{-1}$ the system is linearly stable. It should be noted that while a fluid with $R_0^{-1} > R_c^{-1}$ may be linearly stable, it is still possible for an instability to be triggered through finite amplitude perturbations \citep[assuming that the perturbations are of the right functional form, see][]{huppert1976,proctor1981}. When we discuss ODDC, however, we are referring only to the linearly unstable kind of double-diffusive convection. 

%section: Linear stability analysis
\section{Linear stability analysis} \label{sec:LinStab}
We analyze the linear stability of rotating double-diffusive convection by first linearizing the governing equations in (\ref{eq:GovEq}) around $\tilde{T}=\tilde{\mu}=\mathbf{u}=0$. We then assume that the functional form of the perturbations is
\begin{equation}
\{\mathbf{u},\tilde{T},\tilde{\mu}\} = \{\mathbf{\hat{u}},\hat{T},\hat{\mu}\} \exp\left( ilx + imy + ikz + \lambda t \right) \, ,
\end{equation}
where the hatted quantities are the mode amplitudes, and where $l$, $m$, and $k$ are the wave numbers for the $x$, $y$, and $z$ directions, respectively. By assuming solutions of this form, we get the following dispersion relation:
\begin{eqnarray} \label{eq:DispRel}
\left( \lambda + {\rm Pr}K^2 \right)^2 \left( \lambda + \tau K^2 \right) \left( \lambda + K^2 \right)& \nonumber \\
- \frac{K_H^2}{K^2}{\rm Pr}\left( \lambda + {\rm Pr}K^2 \right) \left[ \left( \lambda + \tau K^2 \right) - R_0^{-1}\left( \lambda + K^2 \right) \right]& \nonumber \\
+ {\rm Ta^*} \frac{\left( m\sin{\theta} + k\cos{\theta} \right)^2}{K^2} \left( \lambda + \tau K^2 \right) \left( \lambda + K^2 \right)& = 0 \, ,
\end{eqnarray}
where $K=\sqrt{l^2+m^2+k^2}$ and $K_H$ is the magnitude of the horizontal wavenumber defined as $K_H=\sqrt{l^2+m^2}$. \citet{Worthem1983} presented a similar linear stability analysis for rotating fingering convection (which has a similar dispersion relation to ODDC) which also included vertical velocity gradients and lateral gradients of temperature and chemical composition. As expected, when the additional physical effects are removed, and when the background gradients of temperature and chemical composition are assumed to be negative, their dispersion relation \citep[Equation 6.2 of][]{Worthem1983} is equivalent to the one shown here. Other similar linear stability analyses of rotating double diffusive systems can be found in \citet{kerr1986} and \citet{kerr1995}.

As in non-rotating ODDC, it can be shown that the fastest growing linear modes in the rotating case have purely vertical fluid motions which span the height of the domain (ie. $k=0$). In fact, in Equation (\ref{eq:DispRel}) we see that when $\theta=0$ and $k=0$ the rotation-dependent term drops out altogether. The fastest growing modes in rotating systems with $\theta=0$ are therefore identical to their non-rotating counterparts both in horizontal wavenumber and growth rate.

\begin{figure}[h] % fig 2
\centering
%\epsscale{0.9}
%\subfloat[]{\includegraphics[scale=0.25]{Ta1.eps} \label{fig:LinStabPhase_a}} \hspace{1em}
%\subfloat[]{\includegraphics[scale=0.25]{Ta100.eps} \label{fig:LinStabPhase_b}} \hspace{1em}
%\subfloat[]{\includegraphics[scale=0.25]{Ta1000.eps} \label{fig:LinStabPhase_c}} \\
%\subfloat[]{\includegraphics[scale=0.30]{theta_025.jpg} \label{fig:LinStabPhase_d}} \hspace{1em}
%\subfloat[]{\includegraphics[scale=0.30]{theta_0375.jpg} \label{fig:LinStabPhase_e}} \hspace{1em}
%\subfloat[]{\includegraphics[scale=0.30]{theta_05.jpg} \label{fig:LinStabPhase_f}}
\includegraphics[width=\linewidth]{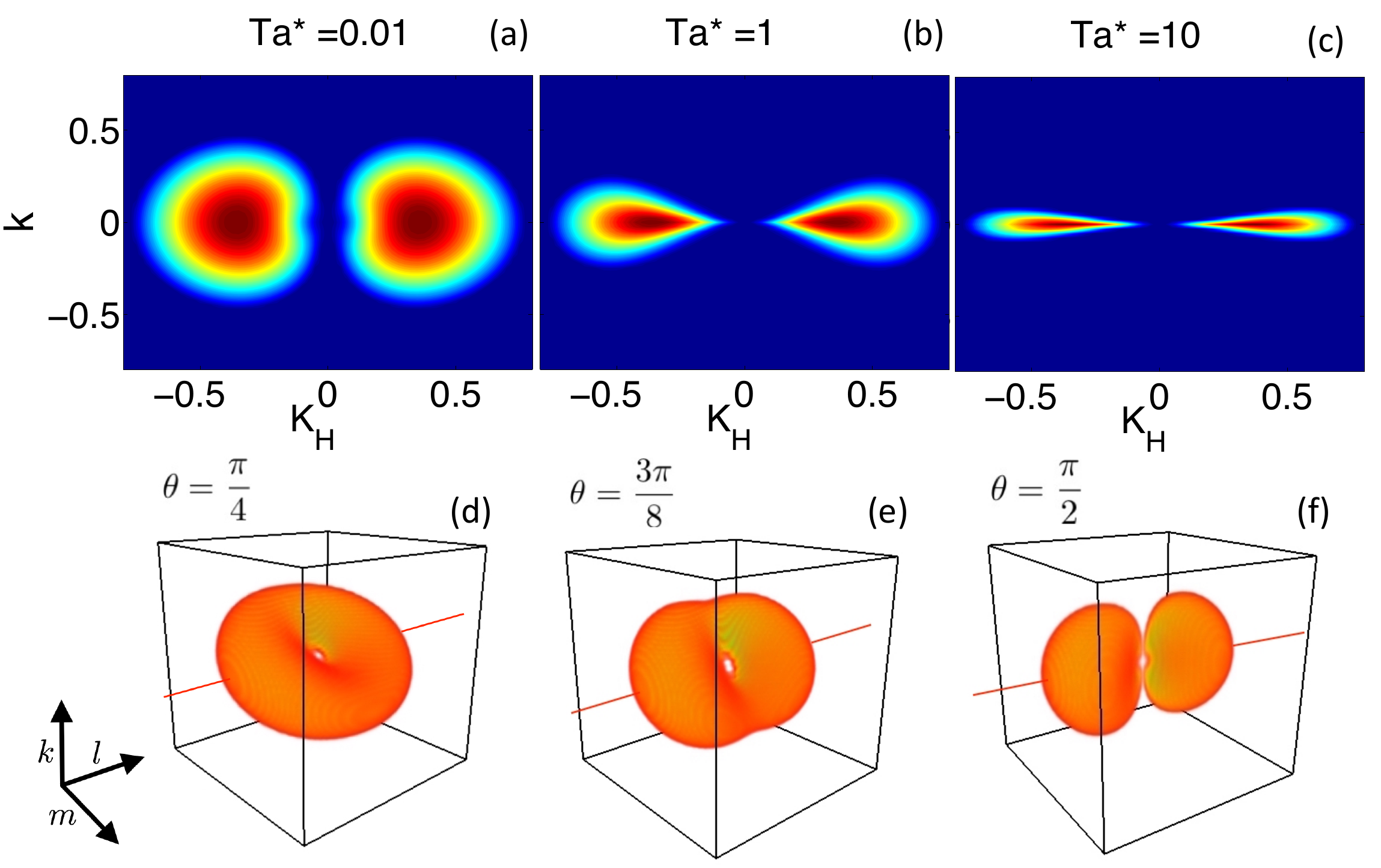}
\caption{In each of the panels, ${\rm Pr}=\tau=0.1$, $R_0^{-1}=1.25$. Panels (a-c): Growth rates versus horizontal and vertical wave numbers for stated values of ${\rm Ta^*}$ with $\theta=0$. Panels (d-f): Surface of null growth rate for ${\rm Ta^*} = 1$ and stated values of $\theta$. The line shows the axis of $l$ wavenumbers. All points on this axis are unaffected by rotation, including the fastest-growing modes. }
 \label{fig:LinStabPhase}
\end{figure}

However, when $\theta = 0$ rotation does affect modes with $k \neq 0$, and always acts to reduce their growth rates. This is illustrated in Figure \ref{fig:LinStabPhase} which shows mode growth rates as a function of $k$ and $K_H$ for various values of ${\rm Ta^*}$. In this ``polar" configuration, the mode growth rate only depends on the total horizontal wavenumber $K_H$, and not on $l$ or $m$ individually. As rotation increases we see that modes with $k \neq 0$ grow more slowly or become stable, while only modes with very low $k$ or $k = 0$ remain unstable.

When $\theta \neq 0$ the fastest growing modes still have $k=0$. However, to avoid the attenuating effect of rotation on their growth rates, they must satisfy the additional constraint, 
\begin{equation} \label{eq:InclinedConstraint}
m\sin{\theta} = -k\cos{\theta} \, . 
\end{equation}
Consequently, the fastest growing modes must have both $m=0$ and $k=0$. Because of this extra constraint, there are fewer modes that grow at the fastest rate. This is well illustrated in Figure \ref{fig:LinStabPhase} where we see that in rotating ODDC there is a ring of modes that are unaffected by rotation that is inclined at an angle of $\theta$. When $\theta \neq 0$, this ring intersects the $k=0$ plane at only two points meaning that there are only two fastest growing modes whose growth rates are not diminished by the effects of rotation. These unaffected fastest growing modes take the form of invariant vertically oscillating planes, spanned by the direction of gravity and the rotation axis (see Section \ref{sec:incSims} for more details on this limit).

% SIMULATION WITH THETA=0
\section{Simulations with $\theta=0$} \label{sec:thetaZero}
Reproducing the conditions of stellar or planetary interiors in laboratory experiments is practically impossible, so in order to understand the development of rotating ODDC beyond linear theory, we must study results from direct numerical simulations (DNS). In this section we analyze data from 3D numerical simulations run using a version of the pseudo-spectral, triply periodic, PADDI Code \citep{traxler2011}, which has been modified to take into account the effects of rotation. Each simulation is run with ${\rm Pr}=\tau=0.1$, $R_0^{-1}=1.25$. We have chosen these values because non-rotating simulations at these parameters have been found to spontaneously form layers \citep[as predicted by $\gamma$-instability theory, see][]{Mirouh2012} which allows us to evaluate how global rotation affects the formation and evolution of these layers. We focus on 5 simulations with ${\rm Ta^*}=0,0.01,0.1,1$ and $10$. Based on their qualitative behavior, we consider the simulations with ${\rm Ta^*}=0.01,$ and $0.1$ to be ``low ${\rm Ta^*}$" and the simulations with ${\rm Ta^*}=1$ and $10$ to be ``high ${\rm Ta^*}$". Each has an effective resolution of $384^3$ mesh points and the simulation domains have dimensions of $(100d)^3$. The simulations are initialized with random infinitesimal perturbations to the temperature field.

When studying the behavior of rotating ODDC using DNS, the quantities of greatest relevance to astrophysical models are the vertical fluxes of temperature and chemical composition through the domain. We express these fluxes in terms of thermal and compositional Nusselt numbers, ${\rm Nu}_T$ and ${\rm Nu}_{\mu}$, which are measures of total fluxes (turbulent $+$ diffusive) in units of the diffusive flux. Using the non-dimensionalization described in Section \ref{sec:mathMod}, ${\rm Nu}_T$ and ${\rm Nu}_{\mu}$ are expressed as
\begin{eqnarray}
{\rm Nu}_T(t) &= &1 + \langle \tilde{w}\tilde{T} \rangle = 1 + \left\langle \left| \nabla \tilde{T} \right|^2 \right\rangle \, , \\
{\rm Nu}_{\mu}(t) &= &1 + \frac{\langle \tilde{w}\tilde{\mu} \rangle}{\tau R_0^{-1}} = 1 +  \frac{\left\langle \left| \nabla \tilde{\mu} \right|^2 \right\rangle}{\left(R_0^{-1}\right)^2} \, ,
\end{eqnarray}
where angle brackets denote an average over all three spatial dimensions, and $\left| \nabla \tilde{T} \right|^2$ and $\left| \nabla \tilde{\mu} \right|^2$ are the thermal and compositional dissipations \citep[for a detailed explanation of the dissipations, see][]{Wood2013,Moll2016}. In practice, we are most interested in the capacity of ODDC to induce vertical turbulent mixing. We therefore quantify transport in terms of the non-dimensional turbulent flux of temperature, ${\rm Nu}_T-1$, and the non-dimensional turbulent flux of chemical species, ${\rm Nu}_{\mu}-1$. These quantities can also be viewed as the ratio of turbulent diffusivity to the microscopic diffusivity for each transported quantity.

We also note that for astrophysical objects it is usually possible to estimate the heat flux by observing the intrinsic luminosity, but direct measurements of the compositional flux are more difficult to obtain. However, the rate of compositional transport may be inferred through observations of the heat flux if a set relationship exists between them. For this reason we also express our results in terms of the total inverse flux ratio, $\gamma_{\rm tot}^{-1}$, given (non-dimensionally) by
\begin{equation}
\gamma_{\rm tot}^{-1} = \tau R_0^{-1} \frac{{\rm Nu}_{\mu}}{{\rm Nu}_T} \, .
\end{equation}
This is the the ratio of the total buoyancy flux due to compositional transport, to the total buoyancy flux due to heat transport, which was first discussed by \citet{stevenson1977}. This ratio is typically smaller than one in the double-diffusive regime when significant turbulent mixing occurs, and describes what fraction of the total energy flux can be used to mix high-$\mu$ chemical species upwards. The inverse flux ratio is also a crucial player in the $\gamma-$instability theory: indeed, as shown by \citet{Mirouh2012}, a necessary and sufficient condition for layer formation in ODDC is that $\gamma_{\rm tot}^{-1}$ be a decreasing function of $R_0^{-1}$. Furthermore, $d \gamma_{\rm tot}^{-1} / dR_0^{-1}$ controls the growth rate of layering modes. 

Finally, measuring how the relative influence of rotation changes as rotating simulations evolve offers insight into how our results may scale to larger systems. We measure the influence of rotation with a Rossby number (the ratio of the inertial force to the Coriolis force), which is usually defined as a turbulent velocity divided by the product of a length scale and the rotation rate. Here, we define the Rossby number as
\begin{equation} \label{eq:RoNum}
{\rm Ro} = \frac{u_{\rm h,rms}}{2 \pi L_h  \sqrt{\rm Ta}^*} \, ,
\end{equation}
where $u_{\rm h,rms}$ is the rms horizontal velocity, and $L_h$ is the expectation value of the horizontal length scale of turbulent eddies over the power spectrum, defined as 
\begin{equation}
L_h = \frac{ \sum_{l,m,k} \frac{\left( |\hat u_{lmk}|^2 + |\hat v_{lmk}|^2 \right)}{\sqrt{l^2 + m^2}} }{ \sum_{l,m,k}  (|\hat u_{lmk}|^2 + |\hat v_{lmk}|^2 ) } \, ,
\end{equation}
where $\hat u_{lmk}$ and $\hat v_{lmk}$ are the amplitudes of the Fourier modes of $u$ and $v$, respectively, with wavenumber $(l,m,k)$. We define ${\rm Ro}$ this way because in systems where $\theta=0$, only horizontal velocity components are affected by rotation.

% subsection: Growth and saturation of the linear instability
\subsection{Growth and saturation of the linear instability}
\begin{figure}[h]%fig 3
\centering
\includegraphics[width=0.5\linewidth]{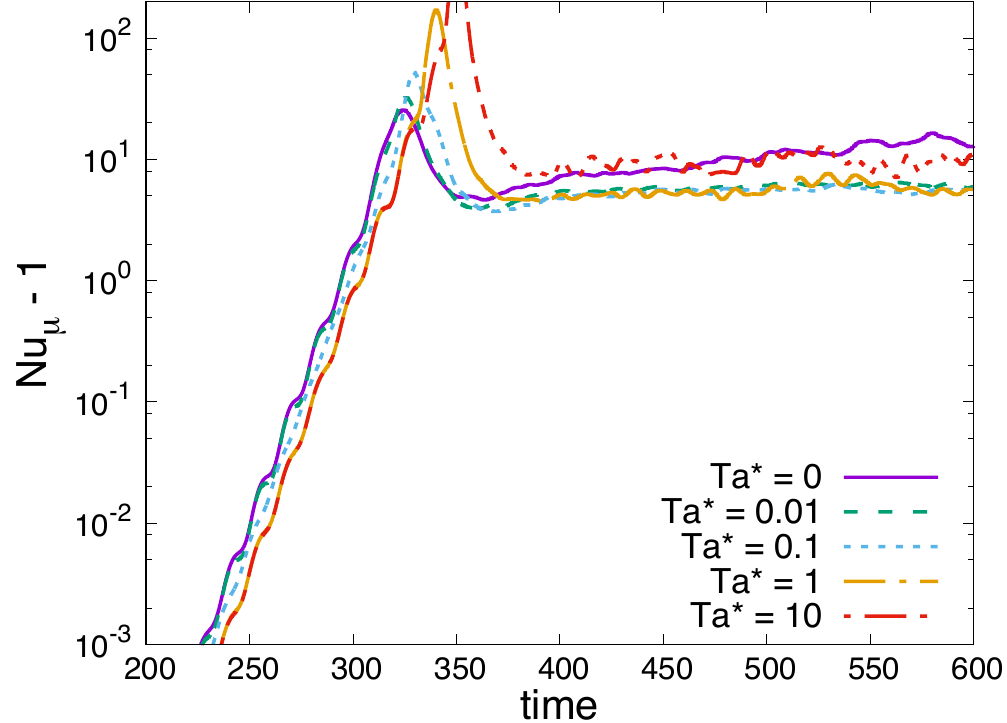}
\caption{Exponential growth and early stages of the non-linear saturation of the turbulent compositional flux for simulations with ${\rm Pr}=\tau=0.1$, $R_0^{-1}=1.25$, $\theta=0$ and stated values of ${\rm Ta^*}$. The growth rates are independent of ${\rm Ta^*}$. The fluxes immediately after non-linear saturation are also more-or-less independent of ${\rm Ta^*}$, except for the cases with ${\rm Ta}^* = 0$ and ${\rm Ta}^* = 10$ (see text for detail).}
 \label{fig:basicInstNoInc}
 \end{figure}

Figure \ref{fig:basicInstNoInc} shows the turbulent compositional flux as a function of time for each of the simulations with ${\rm Pr} = \tau = 0.1$, and $R_0^{-1} = 1.25$, focusing on the growth and saturation of basic instability of ODDC. It clearly shows that the growth of the linear instability in simulations of rotating ODDC (with $\theta = 0$) behaves similarly to the non-rotating case. This is not surprising since the overall growth of the linear instability is dominated by the fastest growing modes, which in this case are completely unaffected by rotation (see Section \ref{sec:LinStab}).

After the initial growth of the linear instability, each simulation reaches a non-linear saturation (at around $t=300$ in each case) and becomes homogeneously turbulent. Figure \ref{fig:basicInstNoInc} shows that the compositional flux in the homogeneously turbulent phase are roughly independent of ${\rm Ta^*}$ at low ${\rm Ta^*}$. For ${\rm Ta^*}=0.01, 0.1$ and $1$ the mean fluxes during this phase are statistically similar to one another, while the most rapidly rotating simulation (${\rm Ta^*}=10$) reaches a plateau that is slightly higher than the others. Note that the composition flux in the non-rotating simulation (${\rm Ta^*}=0$) behaves differently, because in this case layers begin to form almost immediately after saturation. This causes it to continue to grow after saturation (albeit at a slower rate), never achieving a quasi-steady state as the rotating simulations do. We now look in more detail at the behavior of the low Ta* and high Ta* sets of simulations, respectively.

% subsection: Low Ta* simulations
\subsection{Low $\rm Ta^{*}$ simulations} \label{sec:LowTa}
\begin{figure}[h] %fig 4
\centering
%\subfloat[]{\includegraphics[scale=0.90]{fig4a.eps} \label{fig:fig_flux_a}} \\
%\subfloat[]{\includegraphics[scale=0.90]{fig4b.eps} \label{fig:fig_flux_b}}
\includegraphics[width=\linewidth]{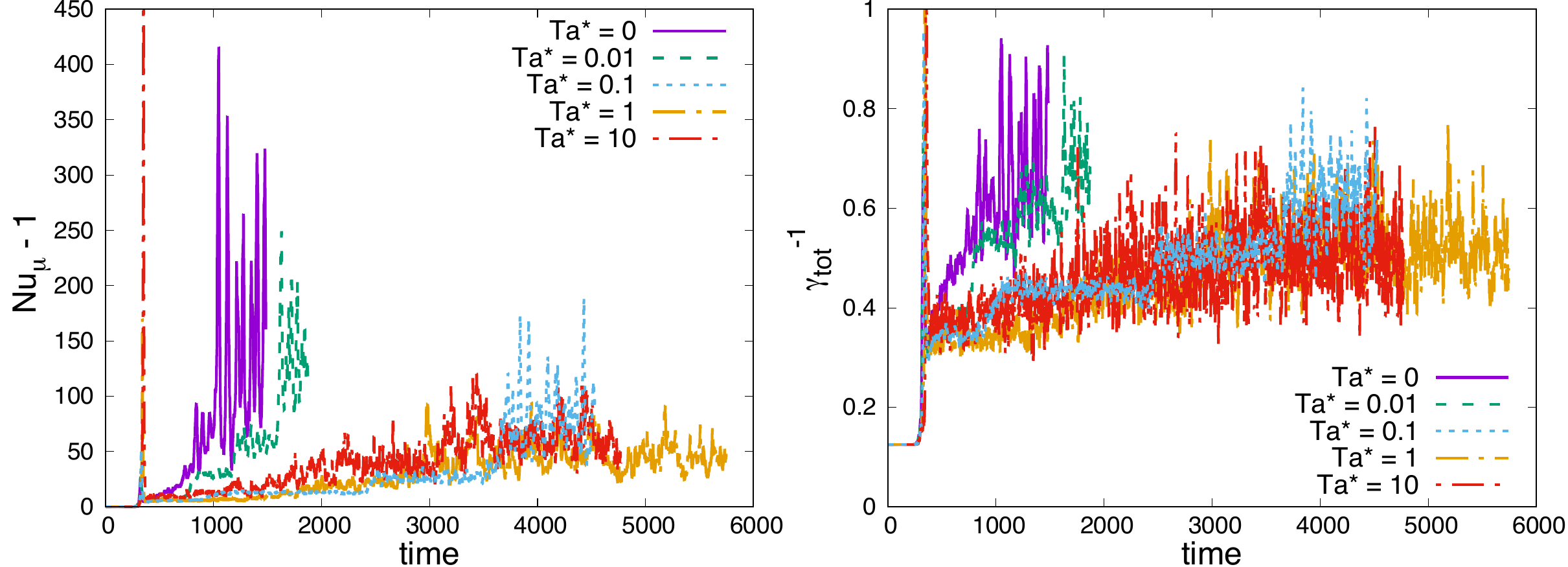}
\caption{Long-term behavior of the turbulent compositional flux (left) and of $\gamma_{\rm tot}^{-1}$ (right) for stated values of ${\rm Ta^*}$. In each simulation, ${\rm Pr}=\tau=0.1$, $R_0^{-1}=1.25$, and $\theta=0$. In low ${\rm Ta^*}$ simulations, the turbulent compositional flux increases in a stepwise manner indicative of layer formation, while in the high ${\rm Ta^*}$ cases there is no clear evidence for similar stepwise increases. }
\label{fig:fig_flux}
\end{figure}

Figure \ref{fig:fig_flux} shows that in low ${\rm Ta^*}$ simulations the homogeneously turbulent phase (where fluxes remain more or less statistically steady) is followed by a series of step-wise increases in the compositional (and thermal) flux, which are indicative of layers that form spontaneously through the $\gamma$-instability and then merge progressively over time. In each case, three layers initially form which then merge into two, and ultimately into a single layer with a single interface. This final configuration is statistically stationary. We therefore find the qualitative evolution of layers to be consistent with previous studies of non-rotating layered ODDC \citep{rosenblum2011,Wood2013}. However, the progressive increase in rotation rate introduces quantitative differences between rotating and non-rotating cases, even at low ${\rm Ta^*}$. The rotation rate clearly affects the time scales for layer formation and layer mergers, respectively, with stronger rotation leading to delays in both processes.

\begin{figure} %fig 5
\centering
%\subfloat[]{\includegraphics[scale=0.90]{fig5a.eps} \label{fig:LayerModes_a}} \\
%\subfloat[]{\includegraphics[scale=0.90]{fig5b.eps} \label{fig:LayerModes_b}}
\includegraphics[width=\linewidth]{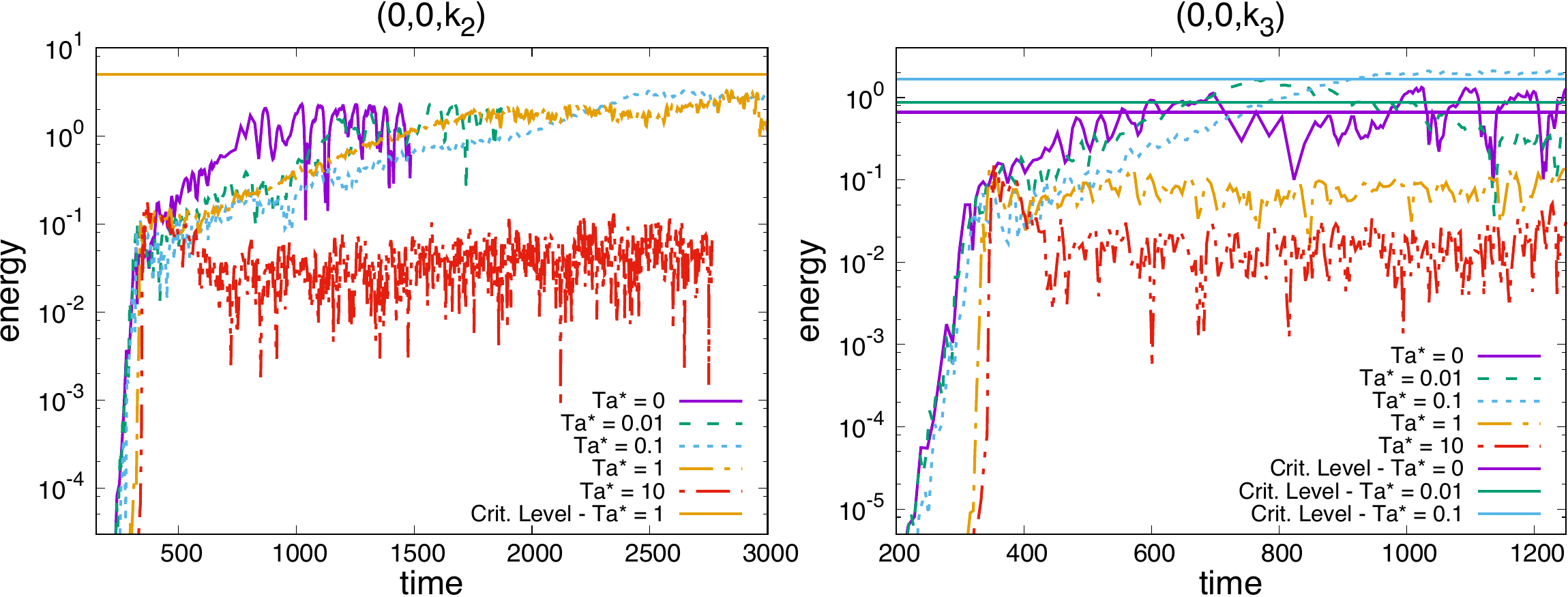}
\caption{Time series of the amount of energy in layering modes $(0,0,k_2)$ (left) and $(0,0,k_3)$ (right) for simulations with ${\rm Pr}=\tau=0.1$, $R_0^{-1} = 1.25$, and $\theta=0$, for various values of ${\rm Ta}^*$. Layering modes are horizontally invariant perturbations to the background profiles of temperature and chemical composition. Also shown are the theoretical amplitudes these layering modes must attain in order to trigger layered convection. Perturbation amplitudes in the low ${\rm Ta^*}$ regime attain this amplitude, but fall short in the ${\rm Ta^*}=1$ simulation. \label{fig:LayerModes}}
\end{figure}

The formation of layers can be understood quantitatively by studying the growth of ``layering modes" predicted by $\gamma$-instability theory \citep[as in][for example]{stellmach2011,rosenblum2011,Mirouh2012}. Each layering mode corresponds to a horizontally invariant, vertically sinusoidal perturbation to the background density profile. To analyze them, we therefore look at the amplitude of the Fourier modes of density perturbations with wave numbers $(0,0,k_n)$, where $k_n=\frac{2 \pi n}{L_z}$, where $L_z = 100d$ is the domain height, and $n$ is the number of layers in the process of forming. The evolution of the $(0,0,k_2)$ and $(0,0,k_3)$ modes as a function of time for each of the three low ${\rm Ta^*}$ simulations is shown in Figure \ref{fig:LayerModes}.

For simulations with ${\rm Ta^*} = 0,0.01$ and $0.1$, $(0,0,k_3)$ is the mode that initially grows to have the largest amplitude, which explains why the staircases first form with three layers. We see that, at low ${\rm Ta^*}$, the $(0,0,k_3)$ modes all initially grow at roughly the same rate. This is unsurprising, since rotation does not have a direct effect on the $\gamma$-instability because Coriolis terms only appear in the momentum equation in (\ref{eq:GovEq}), which is ignored in the mean field theory upon which the $\gamma$-instability is based \citep{Mirouh2012}. Rotation could in principle have an indirect influence over layer formation by significantly affecting the turbulent fluxes in the homogeneously turbulent phase and changing the relationship between $\gamma^{-1}_{\rm tot}$ and $R_0^{-1}$, but as we see in Figure \ref{fig:basicInstNoInc} this is not the case in the low ${\rm Ta^*}$ regime.

To understand the delay in the formation of layers we must instead look at the amplitude that the density perturbations must achieve in order to trigger the onset of layered convection. Indeed, rotation is well-known to delay the onset of instability in the case of thermal convection between parallel plates \citep{chandrasekhar1961}, so by analogy, we expect that the localized positive density gradients caused by the growth of layering modes must be larger to trigger convective overturning and cause the staircase to appear. In the Appendix, we estimate the critical density gradient needed to trigger convection in rotating Rayleigh-B\'enard convection. Using this result, we then compute the amplitude the layering modes must achieve as a function of $k_n$ and ${\rm Ta^*}$, to be
\begin{eqnarray} \label{eq:LayerModeEq}
\left| A_n \right| = \left| \frac{ \frac{3\pi^4}{H_n^4}\left(\frac{H_n^4{\rm Ta}^*}{2{\rm Pr}^2\pi^4}\right)^{\frac{2}{3}} + \frac{27\pi^4}{4H_n^4} + \left( R_0^{-1} - 1 \right) }{2k_n} \right| \, .
\end{eqnarray}
where $H_h = L_z / n = 2\pi/k_n$ is the nondimensional layer height associated with the layering mode $(0,0,k_n)$. This amplitude is shown in Figure \ref{fig:LayerModes} for each of the $(0,0,k_3)$ modes in the low ${\rm Ta^*}$ simulations. Consistent with our idea, the layering modes stop growing shortly after achieving their respective critical amplitudes (except for the Ta* = 1 case, see Section \ref{sec:highTa}). This indicates that layered convection has commenced, taking the form of turbulent convective plumes bounded by freely moving, stably stratified interfaces. 

\begin{figure}[h] %fig 6
\centering
%\subfloat[]{
%\includegraphics[scale=0.8]{fig6a.eps} \label{fig:snapTa0_a} \hspace{3em}
%\includegraphics[scale=0.35]{snapshots/Snap_Ta0.png} \label{fig:snapTa0_b}
%} 
\includegraphics[width=0.9\linewidth]{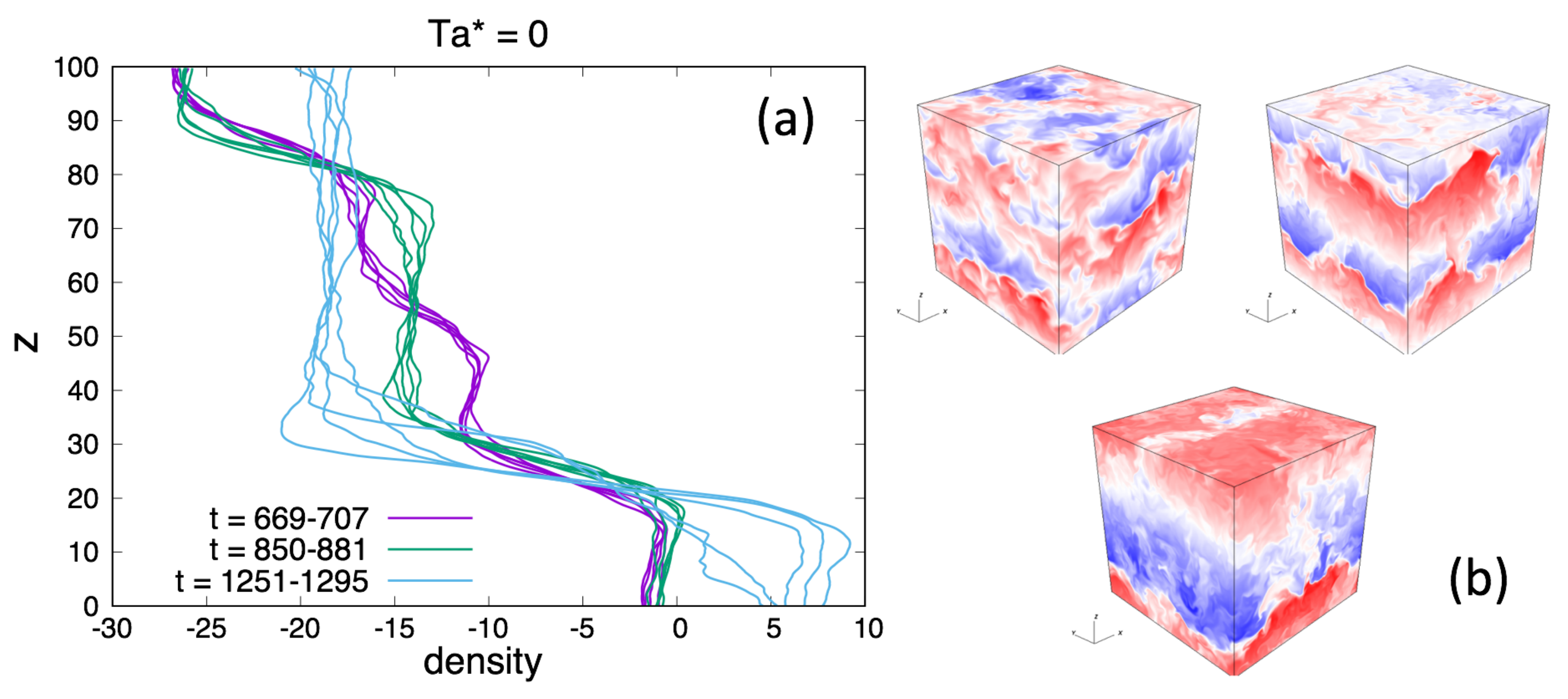}
\caption{(a) Density profiles and (b) snapshots of the chemical composition field in the 3, 2, and 1 layered phases for a non-rotating simulation (${\rm Ta^*}=0$) with ${\rm Pr}=\tau=0.1$ and $R_0^{-1}=1.25$. }
\label{fig:snapTa0}
\end{figure}

\begin{figure}[h] %fig 7
\centering
%\subfloat[]{
%\includegraphics[scale=0.8]{fig7a.eps} \label{fig:snapTa10_a} \hspace{3em}
%\includegraphics[scale=0.35]{snapshots/Snap_Ta10.png} \label{fig:snapTa10_b}
%} 
\includegraphics[width=0.9\linewidth]{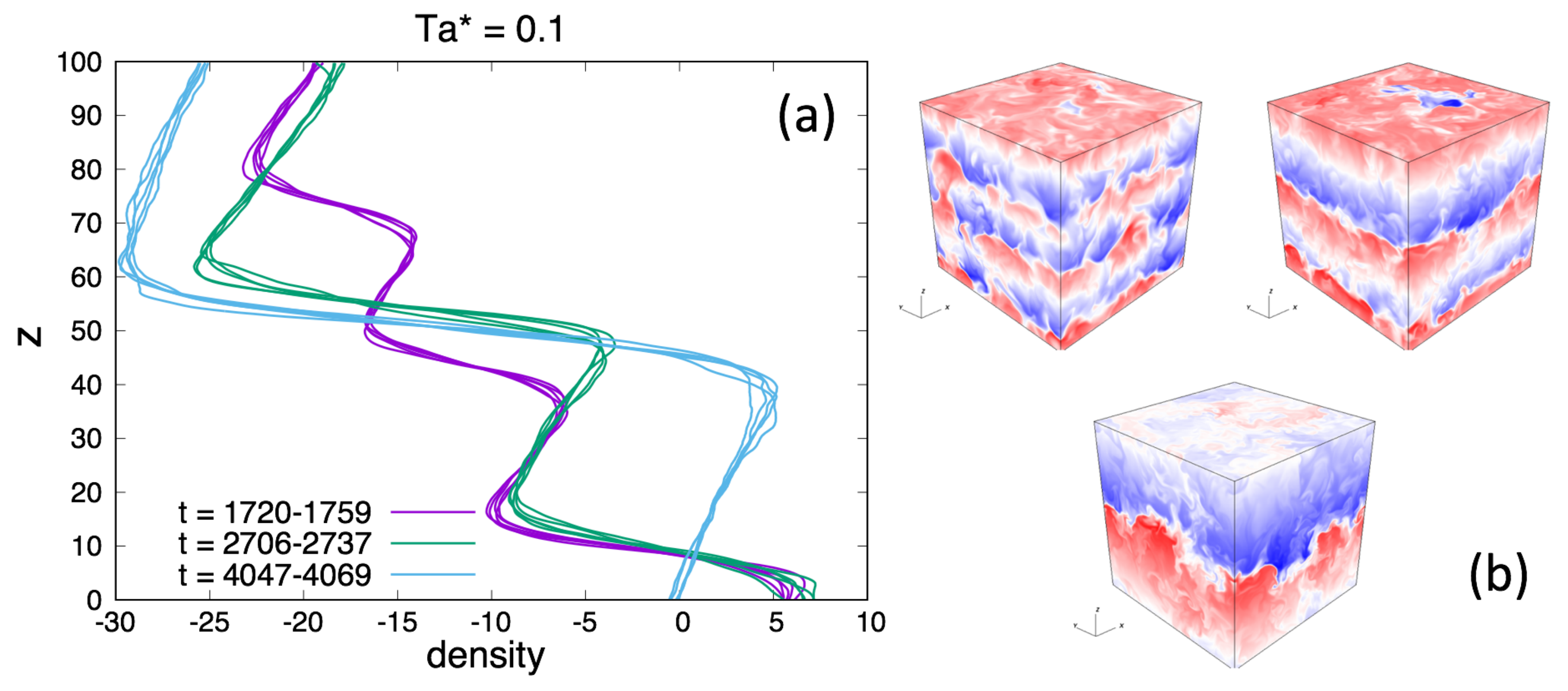}
\caption{(a) Density profiles and (b) snapshots of the chemical composition field in the 3, 2, and 1 layered phases for a simulation with ${\rm Ta^*}=0.1$, ${\rm Pr}=\tau=0.1$, $R_0^{-1}=1.25$ and $\theta=0$. Noteworthy are the layer interfaces which are more stably stratified than in the non-rotating case. Also, there is a larger positive density gradient in the layers themselves.}
\label{fig:snapTa10}
\end{figure}

In each case the mode $(0,0,k_3)$ is then overtaken by modes $(0,0,k_2)$ and ultimately $(0,0,k_1)$ (not shown here). These multi-layer phases are metastable in that they persist over many eddy-turnover times before merging. Snapshots of the 3, 2, and 1-layered phases for ${\rm Ta^*}=0$ and ${\rm Ta^*}=0.1$ are shown in Figure \ref{fig:snapTa0}b and \ref{fig:snapTa10}b.

Rotation also has a strong influence on several aspects of the dynamics of layered convection including the mean density profile within the layers, the stability of the interfaces, and, as mentioned earlier, the merger timescale. In Figures \ref{fig:snapTa0}a and \ref{fig:snapTa10}a, horizontally averaged density profiles show in greater detail the structure of the layers themselves in the three-, two-, and one-layered phases. Stronger rotation is correlated with larger positive density gradients in the layers themselves, which in turn necessarily leads to more stably stratified layer interfaces (at fixed $R_0^{-1}$).

The increase with rotation rate of the density gradients within the layers is similar to what occurs in rotating Rayleigh-B\'enard convection \citep{julien1996}. It is usually argued that turbulent buoyancy mixing by convection adjusts the mean density gradient (outside of any potential boundary layers) to a state of marginal stability. Combined with the fact that the theoretical critical density gradient for marginal stability increases with ${\rm Ta^*}$ (see Equation (\ref{eq:LayerModeEq})), our results are not surprising.

\begin{figure}[h] %fig 8
\centering
\includegraphics[width=0.55\linewidth]{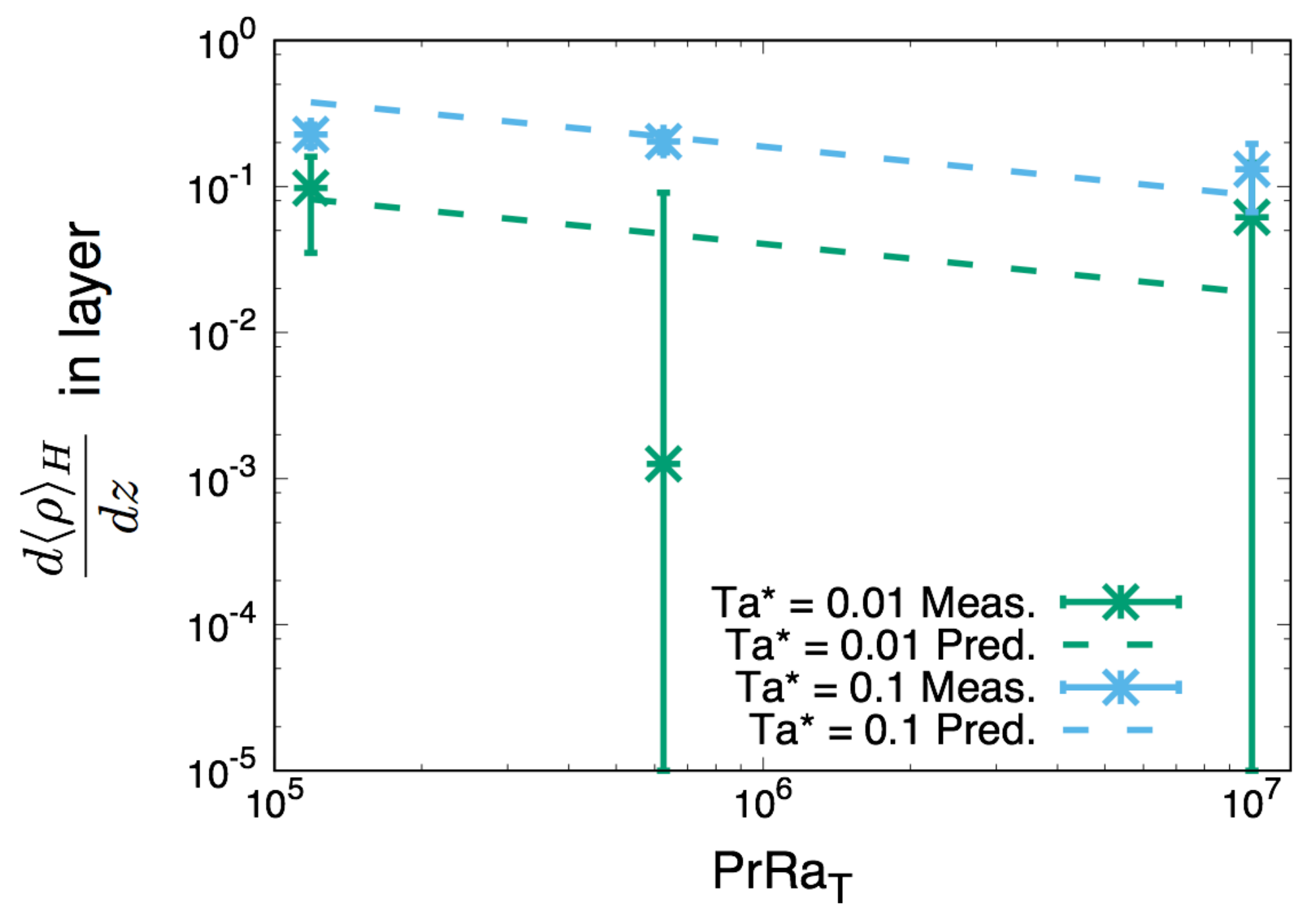}
\caption{Measurements of horizontally averaged density gradients in layers for two low ${\rm Ta^*}$ simulations. Also shown are predicted density gradients for systems with given rotation rates and layer heights. Both simulations have ${\rm Pr}=\tau=0.1$ and $R_0^{-1} = 1.25$.}
\label{fig:DensGrad}
\end{figure}

To study this quantitatively, we first calculate the $z$-derivative of a horizontally averaged density profile and then estimate the gradient in each layer by fitting profiles from a range of time steps over which the layer is stable. We repeat this procedure for the 3, 2, and 1-layered stages in the low $\rm{Ta}^*$ simulations. Figure \ref{fig:DensGrad} shows this data as a function of the product of ${\rm Pr}$ and the thermal Rayleigh number which, in our non-dimensionalization, is defined as
\begin{equation}
{\rm Ra}_T = \frac{g\alpha \left| T_{0z} - T_{0z}^{\rm ad} \right| \left( H_n d \right)^4}{\kappa_T \nu} = H_n^4 \, ,
\end{equation}
Also included are theoretical values of for the density gradient calculated using Equation (\ref{eq:DensGradAp}) in the appendix, and are given by
\begin{equation}
\frac{d\langle \rho \rangle_H}{dz} = \frac{3 \pi^4}{H_n^4} \left( \frac{H_n^4{\rm Ta^*}}{2{\rm Pr}^2\pi^4} \right)^{\frac{2}{3}} + \frac{27\pi^4}{4H_n^4}
\end{equation}
where $\langle \cdot \rangle_H$ represents a horizontal average.

We indeed find that the density gradients in layers are largest for the simulation with the higher rotation rate (${\rm Ta^*}=0.1$), and that the results from this simulation also fit well with prediction. The case with ${\rm Ta^*}=0.01$, on the other hand, shows much greater variability in density gradient, particularly at larger layer heights, making it difficult to determine whether or not the theoretical predictions are valid. From Figure \ref{fig:DensGrad} we also see that in both cases the steepening of the density gradient is somewhat greater for smaller layers. However, this dependence of density gradient on layer height is weaker than the theory suggests.

In Figure \ref{fig:fig_flux} we see that the simulation with ${\rm Ta^*}=0.1$ spends a greater amount of time in the three- and two-layered phases than either of the other two simulations, and consequently takes about twice as long as the non-rotating run (${\rm Ta^*}=0$) to reach the one-layered phase. The root cause is related to the lower supercriticality of convection within the layers combined with the increased stability of the interfaces. Through inspection of the density profiles of our layered simulations, we see that the positions and shapes of the interfaces in rotating simulations have less variability with time compared with non-rotating simulations. We also see in Figure \ref{fig:fig_flux} that the fluxes in the rotating layered systems oscillate less than they do in the non-rotating ones. \citet{Wood2013} already discussed these large amplitude oscillations in non-rotating simulations and attributed them to the presence of large plumes of fluid punching through interfaces periodically causing spikes in transport. In our layered rotating simulations (particularly the ${\rm Ta^*}=0.1$ case) the absence of large amplitude oscillations in the layered phase suggests that the motion of these plumes is inhibited, possibly because convection is weaker and the interfacial gradients are stronger. Since the plumes could be key players in the layer merger events, their suppression in the rotating simulations could also explain why the merger timescale is longer. 

All of these effects also contribute to the reduction of mean fluxes of temperature and chemical composition through density staircases in rotating ODDC compared with the non-rotating case. To show this quantitatively, Table 1 presents measurements of ${\rm Nu}_T-1$ and ${\rm Nu}_{\mu}-1$ (with standard deviations) for each value of ${\rm Ta^*}$. For low ${\rm Ta^*}$ simulations, which clearly form convective layers, the fluxes are measured in the one-layered phase. The simulation with ${\rm Ta^*}=0.01$ shows $2\%$ and $7\%$ decreases in thermal and compositional fluxes, respectively, compared to the non-rotating simulation, while the ${\rm Ta^*}=0.1$ simulation shows $38\%$ and $44\%$ reductions.
\begin{table}
\begin{center}
\begin{tabular}{c | r | c | c | c}
${\rm Ta^*}$ & ${\rm Ta}$ & ${\rm Nu}_T - 1$ & ${\rm Nu}_{\mu} - 1$ & $\gamma_{\rm tot}^{-1}$ \\
\hline
0 & 0 & 25.3 $\pm$ 10.8 & 149.8 $\pm$ 84.4 & 0.68 $\pm$ 0.12 \\
0.01 & 1 & 24.5 $\pm$ 5.2 & 139.0 $\pm$ 37.5 & 0.68 $\pm$ 0.074 \\
0.1 & 10 & 15.5 $\pm$ 4.0 & 82.9 $\pm$ 28.4 & 0.62 $\pm$ 0.067 \\
1 & 100 & 10.2 $\pm$ 2.1 & 46.5 $\pm$ 12.4 & 0.52 $\pm$ 0.061 \\
10 & 1000 & 14.4 $\pm$ 3.1 & 61.7 $\pm$ 14.7 & 0.51 $\pm$ 0.061 \\
10 (narrow) & 1000 & 44.0 $\pm$ 11.2 & 216.5 $\pm$ 58.9 & 0.62 $\pm$ 0.16
\end{tabular}
\caption{Non-dimensional mean turbulent thermal and compositional fluxes through the domain. Mean values are calculated by taking time averages of the turbulent fluxes in the ultimate statistically stationary state achieved by the simulation. In each case, ${\rm Pr}=\tau=0.1$, $R_0^{-1}=1.25$ and $\theta=0$. For the cases with ${\rm Ta}^* = 0$, 0.01 and 0.1, these fluxes are measured in the 1-layered phase. For the cases with ${\rm Ta}^* = 1$ and 10, fluxes are measured once the system reaches a statistically steady state (see Figure \ref{fig:fig_flux}). }
\end{center}
\label{tab:FluxTable}
\end{table} 

\begin{figure}[h] %fig 9
\centering
\includegraphics[width=0.5\linewidth]{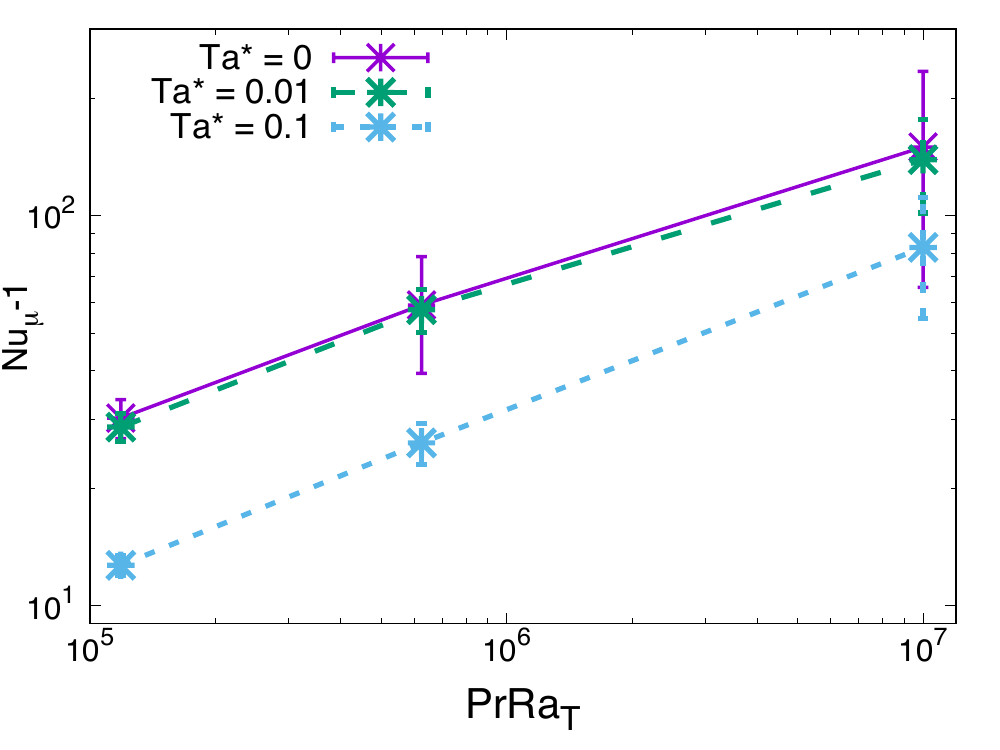}
\label{fig:LayerSats_a}
\caption{Non-dimensional, mean turbulent compositional flux as a function of ${\rm Pr Ra}_T$ for low ${\rm Ta^*}$ simulations with ${\rm Pr}=\tau=0.1$, $R_0^{-1}=1.25$. Mean values are computed by time averaging the instantaneous turbulent compositional flux in the 3-, 2-, and 1-layered phases. In the ${\rm Ta^*}=0.1$ simulations rotation acts to reduce the turbulent compositional flux in each layered phase. However roughly the same power law applies to all simulations. \label{fig:LayerSats}}
\end{figure}

\citet{Wood2013} showed that fluxes in non-rotating layered ODDC follow a power law scaling which depends on the product of the Prandtl number and the thermal Rayleigh number. Figure \ref{fig:LayerSats} shows the mean non-dimensional turbulent compositional flux as a function of ${\rm PrRa}_T$ for each of our low Ta* simulations (${\rm Ta^*}=0$, $0.01$, $0.1$). To collect this data, average fluxes were computed in the 1, 2, and 3 layer phases of each simulation. %As we discuss below in Section \ref{sec:highTa},  whether or not convective layers form in the simulation with ${\rm Ta^*}=1$ is questionable, however we have nonetheless plotted measurements for the fluxes in what could be considered 1 and 2 layer phases based on density profiles.
We clearly see that rotation leads to reduced transport rates in layered convection. This is not entirely surprising because rotation is known to reduce the convective efficiency in Rayleigh-B{\'e}nard convection \citep{Rossby1969}. Bearing in mind the very limited amount of data available, we nevertheless attempt to fit it with flux laws of the form ${\rm Nu}_T-1 = A({\rm PrRa}_T)^a$ and ${\rm Nu}_{\mu}-1 = B({\rm PrRa}_T)^b$, using a nonlinear least square fit. The results are presented in Table 2. 

\begin{figure} %fig 10
\centering
%\subfloat[]{\includegraphics[scale=0.90]{fig9a.eps} \label{fig:RoPlot_a}} \\
%\subfloat[]{\includegraphics[scale=0.90]{fig9b.eps} \label{fig:RoPlot_b}}
\includegraphics[width=\linewidth]{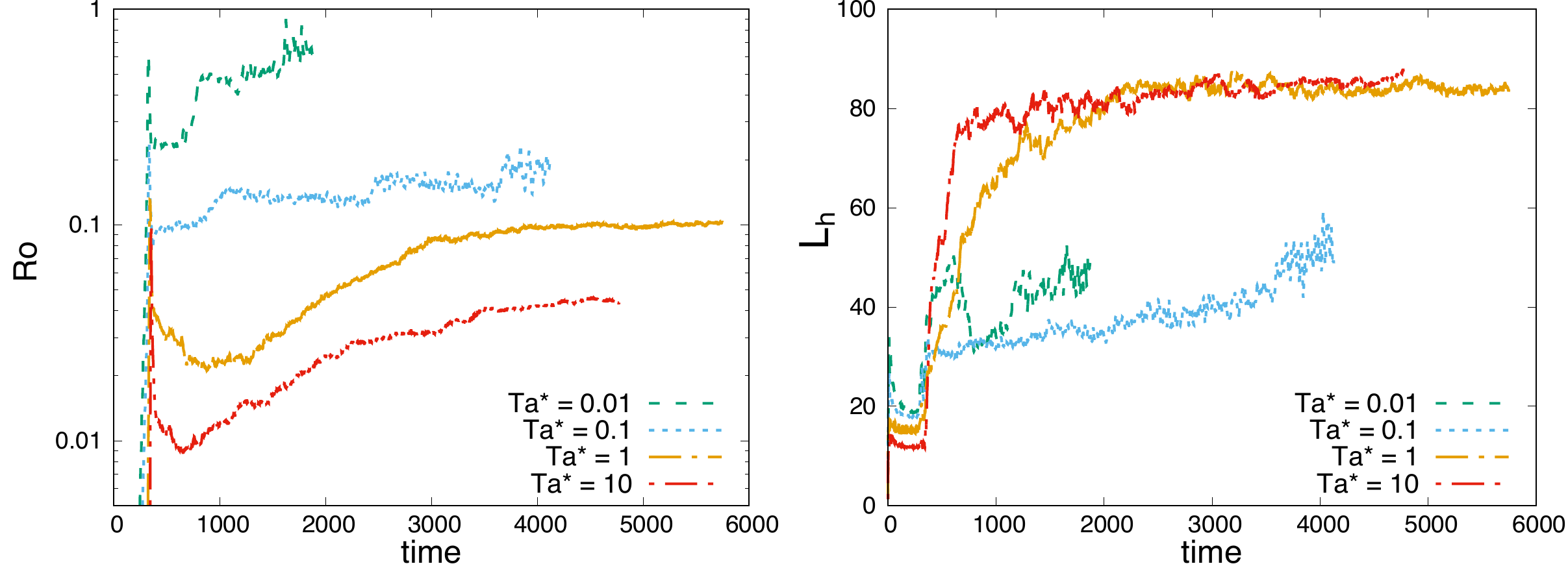}
\caption{Rossby number Ro (left) and average horizontal lengthscale of turbulent eddies $L_h$ (right) for simulations with ${\rm Pr}=\tau=0.1$, $R_0^{-1}=1.25$ and $\theta = 0$, for various values of ${\rm Ta^*}$. Noteworthy is that ${\rm Ro}$ increases as layers merge in the low ${\rm Ta^*}$ regime suggesting a decreased influence of rotation. Also note how the horizontal length scale in the high ${\rm Ta^*}$ simulations, which host a large-scale vortex, is constrained by the domain size.}
\label{fig:RoPlot}
\end{figure}

We find that rotation affects the constants of proportionality ($A$ and $B$), much more than it affects the exponent of ${\rm Pr}{\rm Ra}_T$ ($a$ and $b$). 
For ${\rm Ta^*}=0.01$ rotation has a minimal effect on the fluxes in each layered phase and the relationship between flux and Pr${\rm Ra}_T$ is the same as in the non-rotating case \citep{Wood2013}. For ${\rm Ta^*}=0.1$ however, rotation reduces the coefficient $A$ by almost a factor $5$ and increases the exponent $a$ by around $15\%$. There is evidence however that this change in the exponent may be due to the fact that the relative effect of rotation decreases for increasing values of Pr${\rm Ra}_T$. Indeed, Figure \ref{fig:RoPlot} shows an increase in Rossby number as layers merge in low ${\rm Ta^*}$ simulations (see Equation(\ref{eq:RoNum}) for definition of Rossby number). This suggests that for sufficiently large layer heights rotational effects could become negligible and the flux law probably tends to the one found by \citet{Wood2013}. This will need to be verified in simulations using larger computational domains.
\begin{table} \label{tab:FitData}
\begin{center}
\begin{tabular}{c | r | c | c | c | c} 
\multicolumn{2}{c |}{} & \multicolumn{2}{| c |}{${\rm Nu}_T-1= A({\rm PrRa}_T)^a$} & \multicolumn{2}{| c }{${\rm Nu}_{\mu}-1 = B({\rm PrRa}_T)^b$} \\
\hline
${\rm Ta^*}$ & ${\rm Ta}$ & $A$ & $a$ & $B$ & $b$ \\
\hline
0 & 0 & 0.076 & 0.32 & 0.21 & 0.36 \\
0.01 & 1 & 0.071 & 0.32 & 0.22 & 0.35 \\
0.1 & 10 & 0.016 & 0.37 & 0.035 & 0.42 \\
\end{tabular}
\caption{Best fits for data presented in Figure \ref{fig:LayerSats}. \label{tab:RaTFits}}
\end{center}
\end{table}

% subsection: High Ta* simulations
\subsection{High ${\rm Ta^{*}}$ simulations} \label{sec:highTa}
In contrast to the low ${\rm Ta^*}$ case, the behavior of high ${\rm Ta^*}$ simulations (${\rm Ta^*}=1$ and $10$) is radically different from that described in studies of non-rotating ODDC. In Figure \ref{fig:fig_flux} we see that neither of the high ${\rm Ta^*}$ simulations shows clear stepwise increases in either the compositional flux or $\gamma_{\rm tot}^{-1}$. Instead we see turbulent fluxes that grow slowly and oscillate rapidly after saturation of the linear instability until they reach a highly variable statistically stationary state.

\begin{figure}[h] %fig 11
\centering
%\subfloat[]{
%\includegraphics[scale=0.8]{fig10a.eps} \label{fig:densTa100} \hspace{3em}
%\includegraphics[scale=0.4]{snapshots/Snap_Ta100.png} \label{fig:snapTa100}
%}
\includegraphics[width=0.8\linewidth]{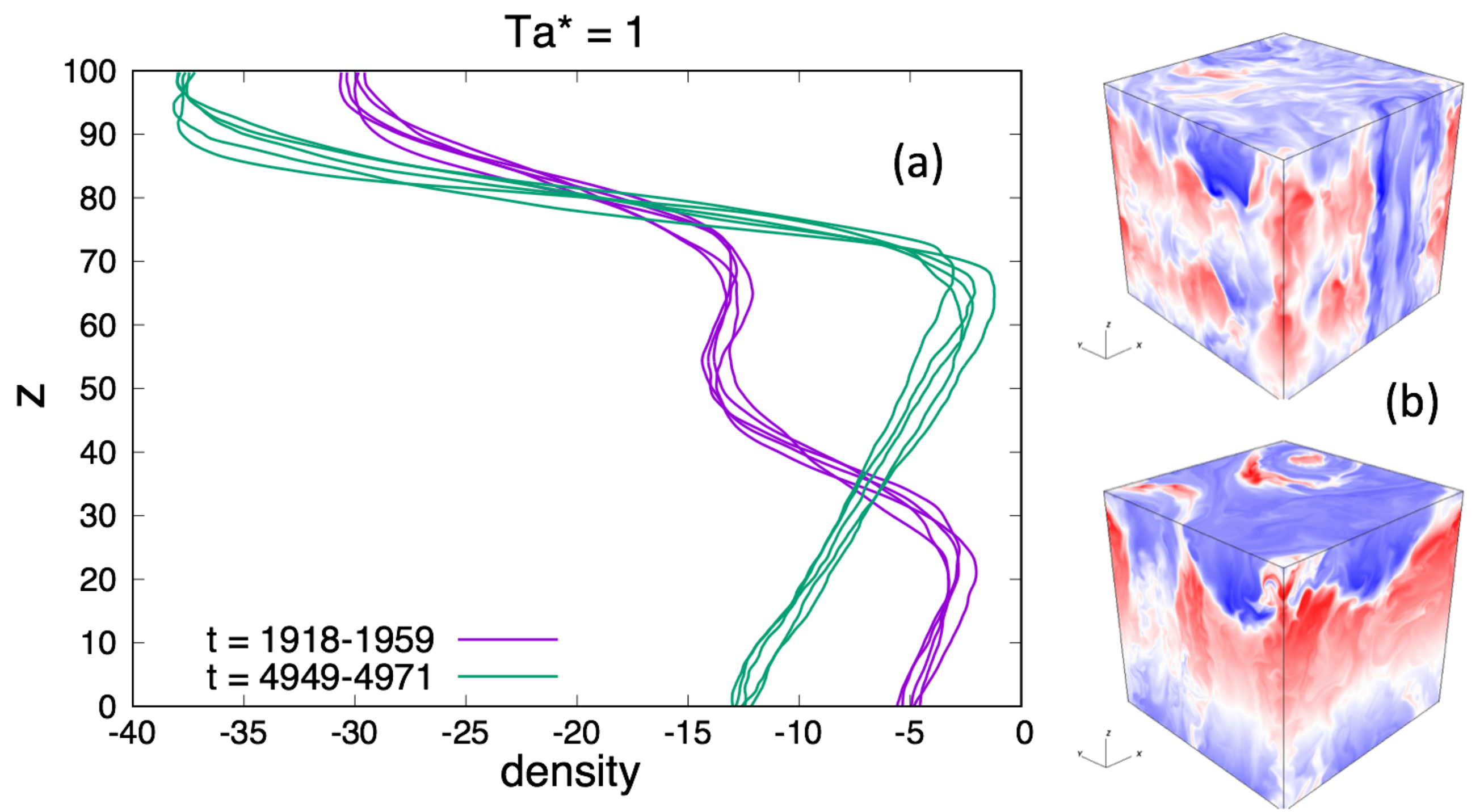}
\caption{(a) Density profiles and (b) snapshots of the chemical composition field for a simulation with ${\rm Ta}^*=1$, ${\rm Pr}=\tau=0.1$, $R_0^{-1}=1.25$ and $\theta=0$. Note the presence of both a large scale vortex and layers. }
\label{fig:snapTa100}
\end{figure}

The Ta* = 1 and Ta* = 10 simulations are themselves quite different from one another. Figure \ref{fig:snapTa100} shows that the growth of step-like density perturbations through the $\gamma$-instability occurs for the ${\rm Ta^*}=1$ simulation just as they do in the low ${\rm Ta^*}$ cases. However, from Figure \ref{fig:LayerModes} we see that their amplitudes never becomes large enough to trigger the onset of convection. The absence of the standard stepwise increase in the fluxes associated with the transition to layered convection also supports the idea that the latter does not happen in this simulation (see Figure \ref{fig:fig_flux}). Interestingly, the snapshot of chemical composition shown in Figure \ref{fig:snapTa100} reveals that the system is dominated by a large scale cyclonic vortex. Inspection of the vertical velocity field shows that it is (roughly) constant within the vortex, which is consistent with Taylor-Proudman balance but is inconsistent with a system composed of convective layers separated by interfaces that resist penetrative motion. In some sense, it is perhaps more appropriate to consider ${\rm Ta^*}=1$ to be a transitional case rather than a high ${\rm Ta^{*}}$ case because it displays features of both high and low ${\rm Ta^{*}}$ regimes.

\begin{figure}[h] %fig 12
\centering
%\subfloat[]{
%\includegraphics[scale=0.8]{fig11a.eps} \label{densTa1000} \hspace{3em}
%\includegraphics[scale=0.25]{snapshots/Snap_Ta1000.png} \label{snapTa1000}
%}
\includegraphics[width=0.9\linewidth]{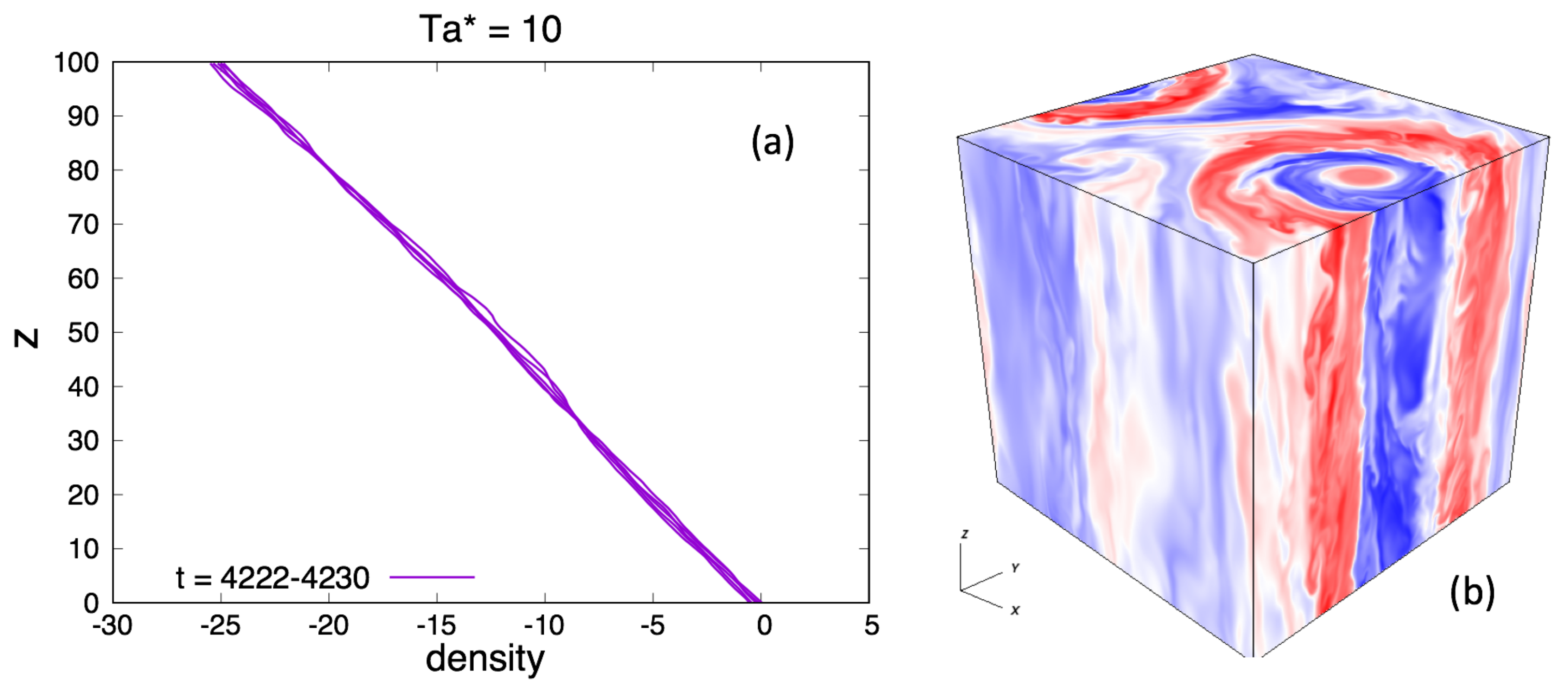}
\caption{(a) Density profiles and (b) snapshots of the chemical composition field for a simulation with ${\rm Ta}^*=10$, ${\rm Pr}=\tau=0.1$, $R_0^{-1}=1.25$ and $\theta=0$. Note the complete absence of perturbations to the background density profile, indicating that layering modes do not grow.}
 \label{fig:snapTa1000}
\end{figure}

\begin{figure}[h] %fig 13
\centering
%\subfloat[]{\includegraphics[scale=0.4]{Ta10_vort.jpg} \label{fig:vorticity_a}} \hspace{3em}
%\subfloat[]{\includegraphics[scale=0.4]{Ta100_vort.jpg} \label{fig:vorticity_b}} \\
%\subfloat[]{\includegraphics[scale=0.4]{Ta1000_vort.jpg} \label{fig:vorticity_c}}
\includegraphics[width=0.9\linewidth]{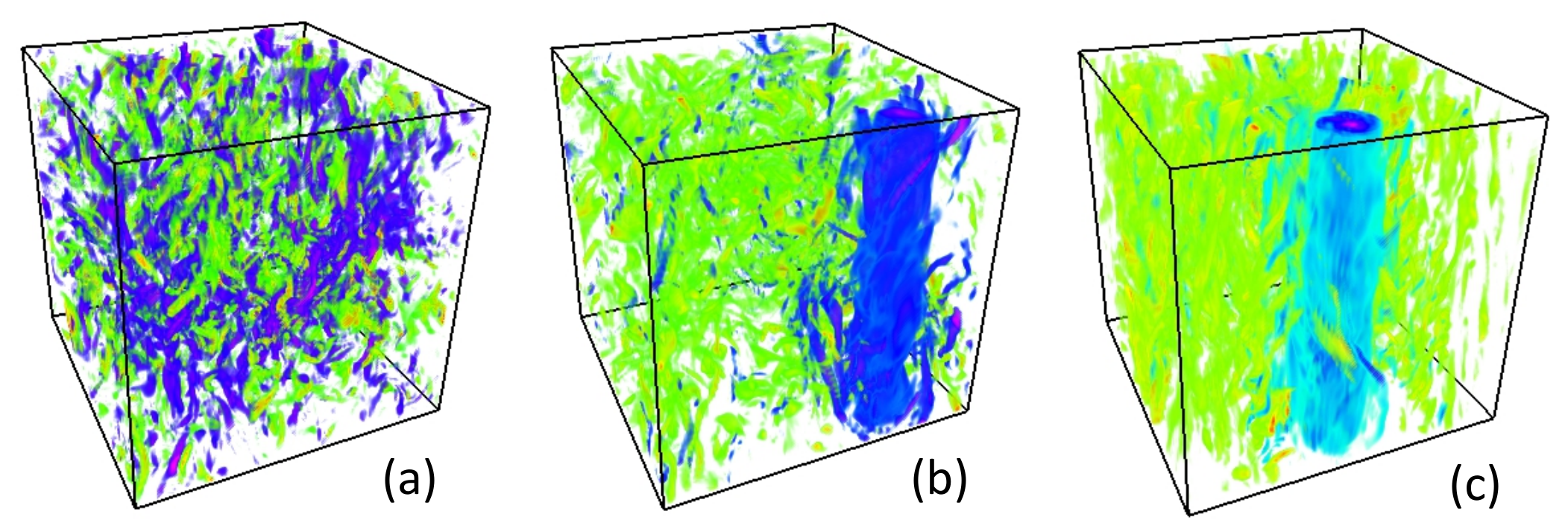}
\caption{Volume-rendered plots of the component of vorticity in the $z$-direction, $\omega_z$, for three simulations with ${\rm Pr}=\tau=0.1$, $R_0^{-1}=1.25$ and $\theta = 0$. (a) ${\rm Ta^*} = 0.1$. (b) ${\rm Ta^*} = 1$. (c) ${\rm Ta^*} = 10$. Purple/blue implies positive (cyclonic) vorticity, while red/yellow implies negative (anticyclonic) vorticity. The first simulation is in the low ${\rm Ta^*}$ regime (${\rm Ta^*}=0.1$) and the other two are in the high ${\rm Ta^*}$ regime (${\rm Ta}^* = 1$ and ${\rm Ta}^* = 10$). Vertically coherent, large scale vortices are present in the high ${\rm Ta^*}$ simulations, but no large-scale coherent structures exist in the low ${\rm Ta^*}$ case.}
\label{fig:vorticity}
\end{figure}

At significantly higher ${\rm Ta^*}$ (in this case ${\rm Ta^*}=10$) we see from Figure \ref{fig:snapTa1000} that the growth of perturbations to the density profile is completely suppressed for the duration of the simulation. Instead, after a transitional period the system becomes dominated by a cyclonic vortex similar to that observed in the ${\rm Ta^*}=1$ simulation albeit with much stronger vorticity.
From the snapshots of vertical vorticity, $\omega_z$, in Figure \ref{fig:vorticity} we see that the simulations with ${\rm Ta^*} = 1$ and $10$ have highly concentrated, vertically invariant vortex cores, necessarily surrounded by a more diffuse region of mostly anti-cyclonic vorticity (since $\int\int \omega_z(x,y,z) dxdy$ = 0 for all $z$). We find that, based on all available simulations, these large-scale vortices only occur in the high ${\rm Ta^*}$ regime. By comparison, the ${\rm Ta^*}=0.1$ simulation shows no large scale coherent structure in the vorticity field (which is true of the other low ${\rm Ta^*}$ simulations as well).
These features are strongly reminiscent of the large scale vortices found by \citet{guervilly2014} in rotating Rayleigh-B\'enard convection using stress-free boundary conditions. In a parameter study they found that Reynolds numbers greater than $300$ and Rossby numbers less than $0.15$ were needed for large scale vortices to form. Using the Reynolds number from \citet{guervilly2014} defined as
\begin{equation}
{\rm Re} = \frac{w_{\rm rms} L_z}{\rm Pr} \, ,
\end{equation}
where $w_{\rm rms}$ is the rms vertical velocity, we find that values of ${\rm Re}$ for our simulations are $\sim 10^3$. Meanwhile, the Rossby numbers are shown in Figure \ref{fig:RoPlot} and are less than $0.1$ for high ${\rm Ta^*}$. This suggests that their vortex formation process may be applicable to our high ${\rm Ta^*}$ simulations despite the significant differences in the systems being studied (ODDC vs. Rayleigh-B\'enard convection). Also as in \citet{guervilly2014}, we find that whenever large scale vortices form they always grow to fill the horizontal extent of the domain\footnote{To verify this, we have run an additional simulation in a domain of horizontal scale $200 d\times  200d$, and height $50d$. The large-scale vortex grew to fill the domain in this case as well.}. \citet{julien2012} proposed that this may always occur in Cartesian domains using the $f-$plane approximation regardless of box size. However they argued that this would be limited in practice by the Rossby radius of deformation in astrophysical objects. Beyond that size, convection or ODDC would likely lead to development of zonal flows in banded structures instead.

\begin{figure}[h] %fig 14
\centering
%\epsscale{0.9}
%\includegraphics[scale=0.9]{fig14.eps}
\includegraphics[width=0.5\linewidth]{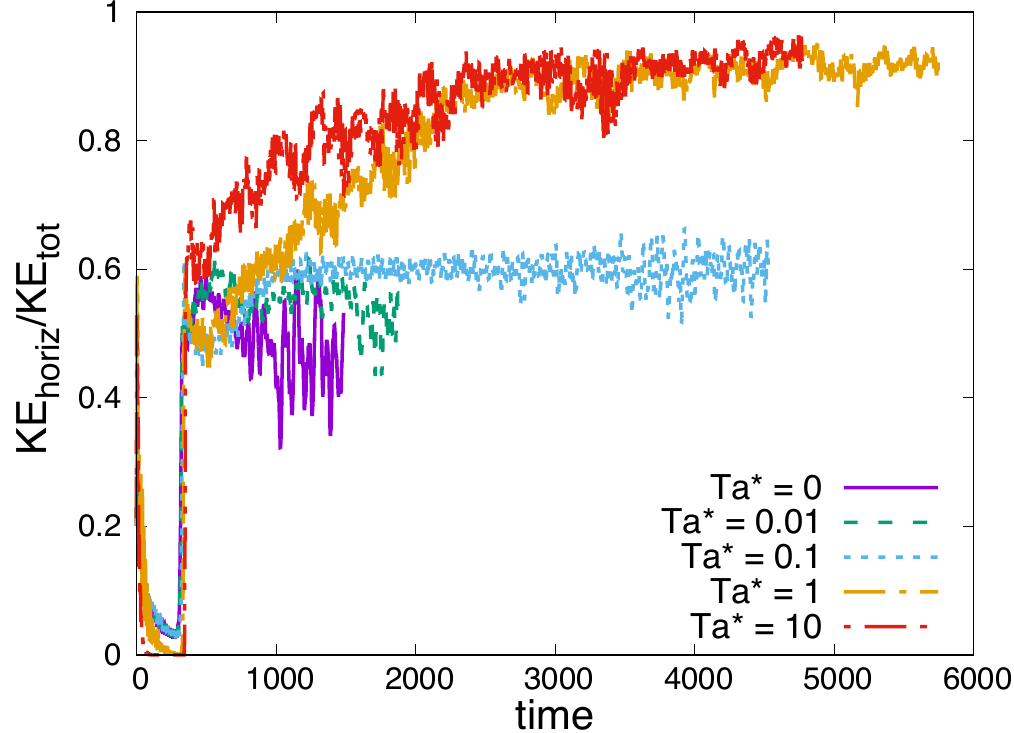}
\caption{Horizontal kinetic energy as a fraction of the total for simulations with ${\rm Pr}=\tau=0.1$, $R_0^{-1}=1.25$, and $\theta = 0$, for various values of ${\rm Ta}^*$. In the high ${\rm Ta^*}$ case this quantity is a proxy for the strength of the large-scale vortices, since they almost entirely dominate the energetics of the system. }
\label{fig:fig_horizKE}
\end{figure}

The amount of energy that can be extracted from ODDC to drive large scale vortices in the high ${\rm Ta^*}$ regime is illustrated in Figure \ref{fig:fig_horizKE} where we see the vast majority of kinetic energy in the ${\rm Ta^*}=1$ and $10$ simulations goes into horizontal fluid motions.
These results showing ratios of horizontal kinetic energy to total kinetic energy are consistent with those calculated in \citet{guervilly2014}. The total amount of kinetic energy in the vertical fluid motions remains the same however in all the simulations, while it is the total kinetic energy of the system that is much larger for high Ta* than for low Ta* simulations.

It is worth noting that while the thermal and compositional fluxes (see Figure \ref{fig:fig_flux}) and vertical velocity stop growing and reach a statistically stationary state (at around $t=3500$ in both high ${\rm Ta^*}$ simulations) the total kinetic energy continues to grow (driven by the continued growth of the horizontal kinetic energy) and has not saturated by $t=$6000. This can be attributed to the fact that horizontal fluid motions are only limited by viscosity, and may only saturate on the global viscous diffusion timescale which is $\sim 10^5$ in these simulations.

Unlike in the low ${\rm Ta^*}$ regime where average fluxes are calculated through time-integration of the quasi-steady 1-layered phase, we choose ranges for time integration of fluxes from $t=3500$ to the end of the simulations in the high ${\rm Ta^*}$ regime. From Table 1 we see that the simulation with ${\rm Ta^*}=1$ has the weakest transport in either ${\rm Ta^*}$ regime with $58.1\%$ and $67.4\%$ reductions to thermal and compositional transport, respectively, compared to the non-rotating case. Interestingly, the ${\rm Ta^*}=10$ simulation shows a slight increases in flux over the ${\rm Ta^*}=1$ case. A possible explanation for this is that the presence of a stably stratified interface separating \st{non-convective} layers in the ${\rm Ta^*}=1$ simulation inhibits vertical motion through the large-scale vortex. This could also suggest that in the high ${\rm Ta^*}$ regime, increased rotation may actually serve to enhance transport rather than suppress it, through vertical motions whose coherence is strengthened by the vortex. However, by contrast with the layered regime, fluxes in the presence of a large scale vortex are highly dependent on the aspect ratio of the box. In Table 1 the narrower simulation at high ${\rm Ta^*}=10$ has significantly higher fluxes than its wider counterpart. This makes it challenging to scale our results to simulations with larger domains, let alone apply them to more realistic astrophysical situations.

Also noteworthy in Table 1 is that $\gamma_{\rm tot}^{-1}$ in the ultimate statistically steady state is roughly the same across all simulations (namely $\gamma_{\rm tot}^{-1} \approx 0.5-0.65$) with high ${\rm Ta^*}$ simulations having a $\gamma_{\rm tot}^{-1}$ that is at most $15\%$ lower than in the low ${\rm Ta^*}$ regime. This is significantly less than the variability that occurs due to the formation of large scale structures (layers or vortices): Figure \ref{fig:fig_flux}b shows how $\gamma_{\rm tot}^{-1}$ increases from roughly $0.35$ in the homogeneously turbulent phase to about $0.6$ in the ultimate stages.

%VARYING PR, TAU, AND R_0^-1
\section{Varying ${\rm Pr}$, $\tau$, and $R_0^{-1}$} \label{sec:diffParams}

We now study the effect of varying ${\rm Pr}$, $\tau$, and $R_0^{-1}$ on both the quantitative and qualitative attributes of rotating ODDC discussed in the previous section. This is not meant to be an exhaustive study, but rather to test whether the conclusions from previous section still hold.

% subsection: Varying Pr and tau
\subsection{Varying ${\rm Pr}$ and $\tau$} \label{sec:varyPrTau}

\begin{figure}[h] %fig 15
\centering
%\subfloat[]{\includegraphics[scale=0.90]{fig14a.eps} \label{fig:fig03_flux_a}} \\
%\subfloat[]{\includegraphics[scale=0.4]{vort_03_1000.jpg} \label{fig:vorticity03_b}}
\includegraphics[width=\linewidth]{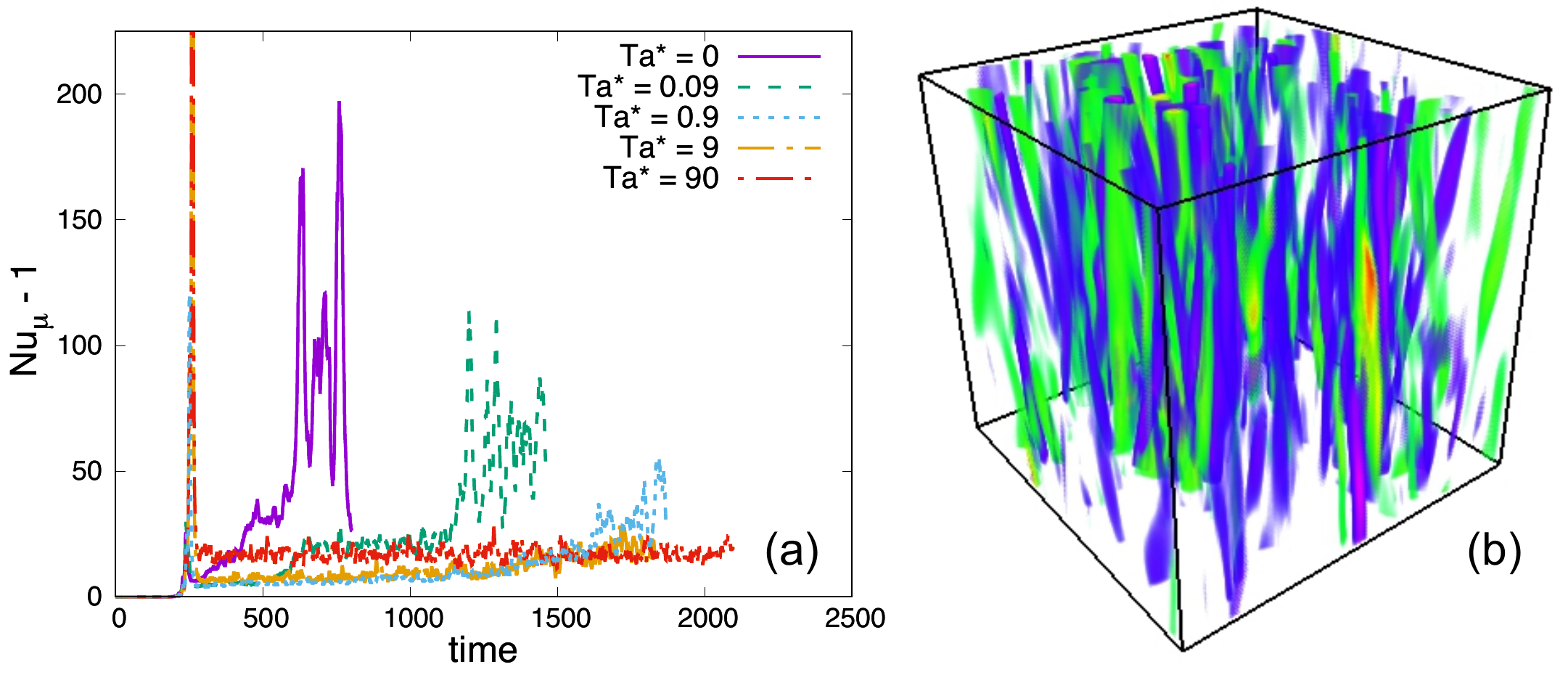}
\caption{(a) Non-dimensional turbulent compositional flux for simulations with ${\rm Pr}=\tau=0.3$ and $R_0^{-1}=1.1$. One simulation is in the low ${\rm Ta^*}$ regime (${\rm Ta^*}=0.3$) and the other three are in the high ${\rm Ta^*}$ regime. (b) Snapshot of the component of vorticity in the $z$-direction for the most rapidly rotating simulation at ${\rm Ta^*}=90$, which appears to be dominated by small scale vortices. This may suggest that large-scale vortices only occur in a specific range of values of ${\rm Ta^*}$ (with $\theta=0$).\label{fig:vorticity03}}
\end{figure}

In order to study the effect of varying ${\rm Pr}$ and $\tau$ we show a set of simulations with ${\rm Pr} = \tau = 0.3$, and $R_0^{-1}=1.1$. As in Section \ref{sec:thetaZero}, we have chosen parameters at which layers form in non-rotating ODDC. Figure \ref{fig:vorticity03} shows the evolution of the turbulent compositional flux as a function of time for simulations with ${\rm Ta^*}=0,0.09,0.9,9,$ and $90$ (corresponding to ${\rm Ta}=0,1,10,100$ and $1000$). Consistent with Section \ref{sec:thetaZero} we find that at low ${\rm Ta^*}$ stepwise increases in mixing rates indicate the transition to layered convection, whereas in the high ${\rm Ta^*}$ case we do not. The transition between low and high ${\rm Ta^*}$ is still ${\rm Ta^*} \approx 1$ (equivalently ${\rm Ta}=10$ when ${\rm Pr}=0.3$). This shows that ${\rm Ta^*}$ is a more appropriate bifurcation parameter than ${\rm Ta}$ to determine when ODDC is rotationally dominated. As in Section \ref{sec:thetaZero}, layer formation only occurs in the low ${\rm Ta^*}$ regime. As with the ${\rm Ta^*}=1$ simulation from Section \ref{sec:thetaZero} which shows characteristics of both high and low ${\rm Ta^*}$ ODDC, the ${\rm Ta^*}=0.9$ simulation here develops a large scale vortex, as well as layer-like perturbations to the background density profile (without evidence of actual layered convection).  

Large scale vortices are observed in simulations with ${\rm Ta^*}=0.9$ and $9$ and look very similar to the corresponding snapshots of the ${\rm Ta^*}=1$ and $10$ simulations in Figures \ref{fig:snapTa100} and \ref{fig:snapTa1000} from the previous section. Interestingly, however the large scale vortex does not form in our most rapidly rotating simulation with ${\rm Ta^*}=90$; instead we see multiple small scale vortices (see Figure \ref{fig:vorticity03}). Analysis of ${\rm Re}$ and ${\rm Ro}$ for this simulation places it in a regime where large scale vortices should form according the criteria of \citet{guervilly2014}. This suggests that there may be additional constraints on the formation of large scale vortices in ODDC which should be determined through a more in-depth survey of parameter space in a future study. Surprisingly, the compositional fluxes for the ${\rm Ta^*}=9$ and $90$ runs are similar, which is likely a coincidence as we saw that the fluxes in the presence of large scale vortices depend on domain size.

% subsection: Simulations at large R_0^-1
\subsection{Simulations at large $R_0^{-1}$}
The simulations we have presented so far were runs with small values of $R_0^{-1}$ which are conducive to layer formation in non-rotating ODDC. However, there is a range of larger values of $R_0^{-1}$ where a system is unstable to ODDC, but where layers are not predicted to spontaneously form through the $\gamma$-instability. Previous studies have shown that, without exception, simulations in this parameter regime remain non-layered for as long as they are run. These simulations are dominated by large scale gravity waves and were studied in depth by \citet{Moll2016} in the context of non-rotating ODDC. In that work they found that the growth of large scale gravity waves is associated with very moderate (but still non-zero) increases in thermal and compositional transport. However, these increases are very small compared to the increases in turbulent transport due to layers, and are likely unimportant for the purposes of stellar and planetary modeling. As a result, turbulent transport by ODDC at $R_0^{-1}$ greater than the layering threshold $R_L^{-1}$ can be ignored.

We now address how rotation affects non-layered ODDC (ie. ODDC at $R_0^{-1}  > R_L^{-1}$). As in Section \ref{sec:thetaZero}, we present five simulations with ${\rm Ta^*} = 0,0.01,0.1,1$ and $10$, and with ${\rm Pr}=\tau=0.1$ and $\theta = 0$. However, for each of these simulations we now set $R_0^{-1} = 4.25$. For comparison, for the stated values of ${\rm Pr}$ and $\tau$, $R_L^{-1} \simeq 1.7$ and the critical inverse density ratio for marginal stability is $R_c^{-1}=5.5$. 

\begin{figure}[h] %fig 16
\centering
%\subfloat[]{\includegraphics[scale=0.2]{gravWv_Ta0.png} \label{fig:snaps425_a}} \hspace{3em}
%\subfloat[]{\includegraphics[scale=0.2]{gravWv_Ta1.png} \label{fig:snaps425_b}} \\
%\subfloat[]{\includegraphics[scale=0.2]{gravWv_Ta10.png} \label{fig:snaps425_c}} \hspace{3em}
%\subfloat[]{\includegraphics[scale=0.2]{gravWv_Ta100.png} \label{fig:snaps425_d}} \\
%\subfloat[]{\includegraphics[scale=0.2]{gravWv_Ta1000.png} \label{fig:snaps425_e}}
\includegraphics[width=0.25\linewidth]{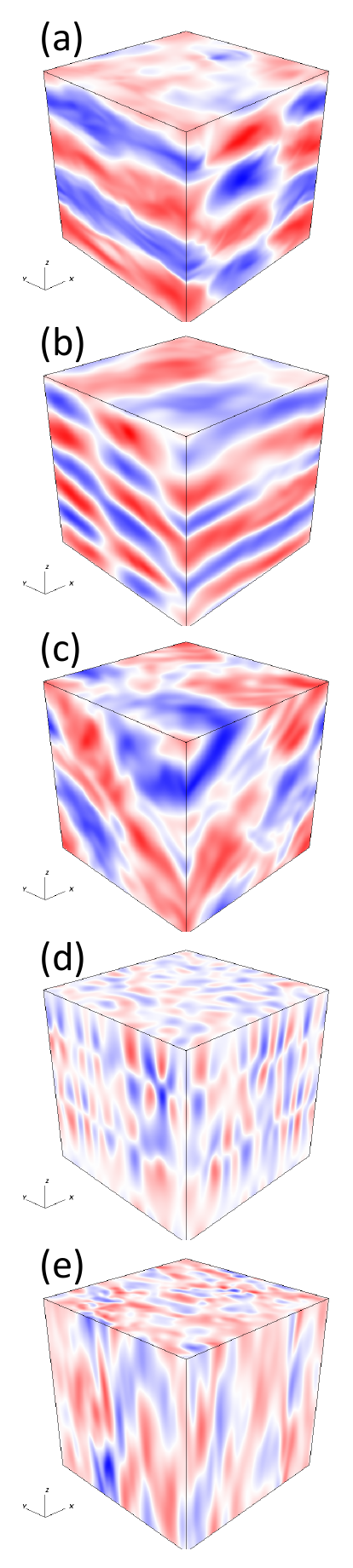}
\caption{Snapshots of the horizontal velocity field ($u$ or $v$) for simulations with ${\rm Pr}=\tau=0.1$, $R_0^{-1}=4.25$, and $\theta = 0$, for various values of ${\rm Ta^*}$: (a) ${\rm Ta}^* = 0$, (b) ${\rm Ta}^* = 0.01$, (c) ${\rm Ta}^* = 0.1$, (d) ${\rm Ta}^* = 1$, and (e) ${\rm Ta}^* = 10$. }
\label{fig:snaps425}
\end{figure}

\begin{figure}[h] %fig 17
\centering
%\subfloat[]{\includegraphics[scale=0.90]{fig16a.eps} \label{fig:fig01_flux_a}} \\
%\subfloat[]{\includegraphics[scale=0.90]{fig16b.eps} \label{fig:fig01_Ro_b}}
\includegraphics[width=\linewidth]{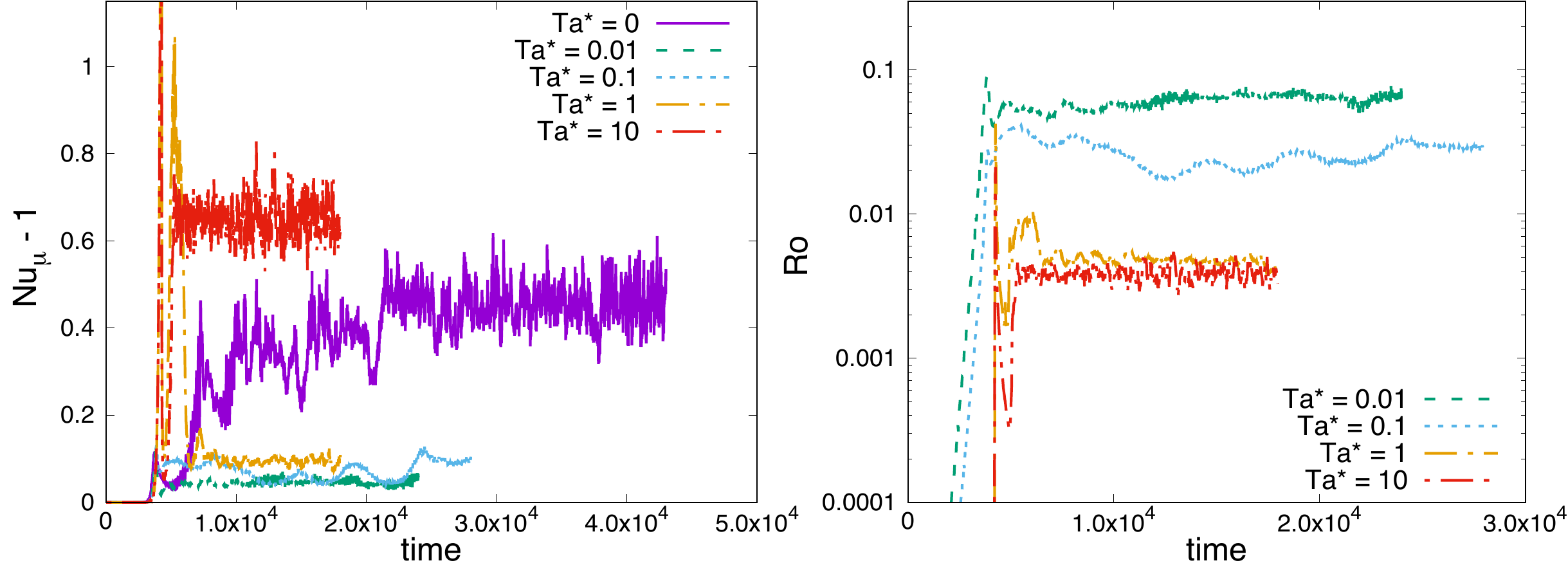}
\caption{Time series of the turbulent compositional flux (left) and Rossby number (right) for simulations with ${\rm Pr}=\tau=0.1$, $R_0^{-1}=4.25$, and $\theta = 0$ for various values of ${\rm Ta^*}$. }
\label{fig:fig01_flux}
\end{figure}

As in Section \ref{sec:thetaZero} we find that high $R_0^{-1}$ simulations can be divided into two general classes of behavior depending on ${\rm Ta^*}$. As seen in the snapshots in Figure \ref{fig:snaps425}, low ${\rm Ta^*}$ simulations are qualitatively similar to non-rotating simulations in that they are dominated by large scale gravity waves. The strongest gravity wave mode in both the ${\rm Ta^*}=0$ and ${\rm Ta^*}=0.01$ simulations has three wavelengths in the vertical direction, one wavelength in the $x$-direction, and is invariant in the $y$-direction. The simulation with ${\rm Ta^*}=0.1$ by contrast is dominated by a larger scale mode with a single wavelength in each spatial direction. Despite their qualitative similarity, inspection of the compositional flux in Figure \ref{fig:fig01_flux} shows large reductions compared to the non-rotating simulation (${\rm Ta^*}=0$), even in the case where ${\rm Ta^*}=0.01$. To understand why this is the case note how ${\rm Ro}$ is small even in the lowest ${\rm Ta^*}$ simulation. This is because the rms velocities are very small in this regime. Rotation therefore plays a role in the saturation of the gravity waves and acts to reduce their amplitudes, which in turn significantly reduces the mixing rates.

As seen in the snapshot in Figure \ref{fig:snaps425}e, the ${\rm Ta^*}=10$ simulation is dominated by vertically invariant vortices, while the ${\rm Ta^*}=1$ is again a transitional case which shows evidence both of gravity waves and of vortices. A significant difference with the results of Section \ref{sec:thetaZero} however, is that vortices at low $R_0^{-1}$ are large-scale, while those at high $R_0^{-1}$ are small-scale (for the same values of ${\rm Ta^*}$). This suggests that the formation of large-scale vortices requires a more unstable stratification (which leads to more turbulence) than is present in the high $R_0^{-1}$ simulations shown here. This is, again, qualitatively consistent with the findings of \citet{guervilly2014} that large scale vortices only form for sufficiently high Reynolds number. %This will be discussed in more detail in Section \ref{sec:conclusion}.

The most rapidly rotating simulation (${\rm Ta^*}=10$) shows a slight increase in the compositional flux compared to the non-rotating simulation but remains far less efficient than layered convection. Importantly, as with the non-rotating simulation, layers never form at any point (when $R_0^{-1} = 4.25$). Consequently, fluxes through non-layered (high $R_0^{-1}$) systems are effectively diffusive, and the conclusion from \citet{Moll2016}, namely that turbulent fluxes are negligible for non-layered systems, remains valid for all the simulations presented here.

%INCLINED SIMULATIONS
\section{Inclined simulations} \label{sec:incSims}
So far, for simplicity, we have discussed simulations in which the rotation vector is aligned with the direction of gravity, and which only model conditions applicable to the polar regions of a star or giant planet. We now discuss the dynamics of ODDC at lower latitudes (ie. simulations with $\theta \ne 0$). In what follows, we return to the parameters studied in Section \ref{sec:thetaZero} (ie. ${\rm Pr}=\tau=0.1$ and $R_0^{-1}=1.25$). We focus on two sets of simulations with ${\rm Ta^*} = 0.1$ and $10$, respectively, which are each comprised of runs with angles $\theta = \frac{\pi}{8}$, $\frac{\pi}{4}$, $\frac{3\pi}{8}$, and $\frac{\pi}{2}$. 

\begin{figure}[h] %fig 18
\centering
\includegraphics[width=0.5\linewidth]{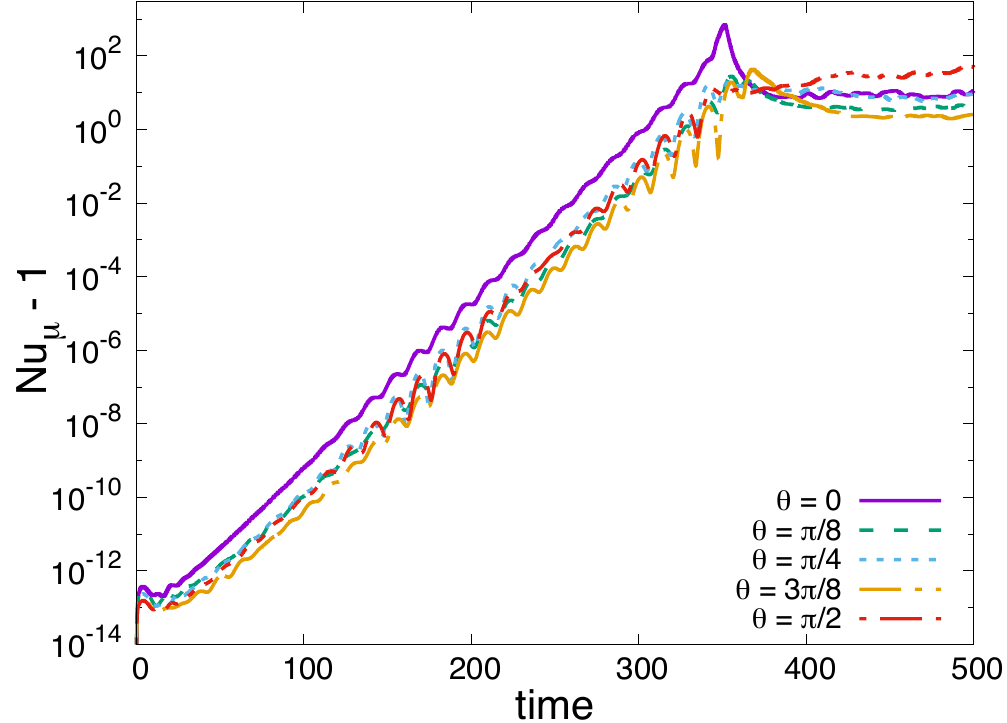}
\caption{Nondimensional turbulent compositional flux during the primary instability growth phase, and immediately following non-linear saturation for simulations with ${\rm Pr}=\tau=0.1$, $R_0^{-1}=1.25$, ${\rm Ta^*}=10$, and various values of $\theta$. }
\label{fig:basicInstInc}
\end{figure}

\begin{figure} %fig 19
\centering
%\subfloat[]{\includegraphics[scale=0.22]{InitGrowth00.png} \label{fig:InstabSnap_a}} \hspace{1em}
%\subfloat[]{\includegraphics[scale=0.22]{InitGrowth0125.png} \label{fig:InstabSnap_b}} \\
%\subfloat[]{\includegraphics[scale=0.22]{InitGrowth025.png} \label{fig:InstabSnap_c}} \hspace{1em}
%\subfloat[]{\includegraphics[scale=0.22]{InitGrowth0375.png} \label{fig:InstabSnap_d}} \\
%\subfloat[]{\includegraphics[scale=0.22]{InitGrowth05.png} \label{fig:InstabSnap_e}}
\includegraphics[width=0.25\linewidth]{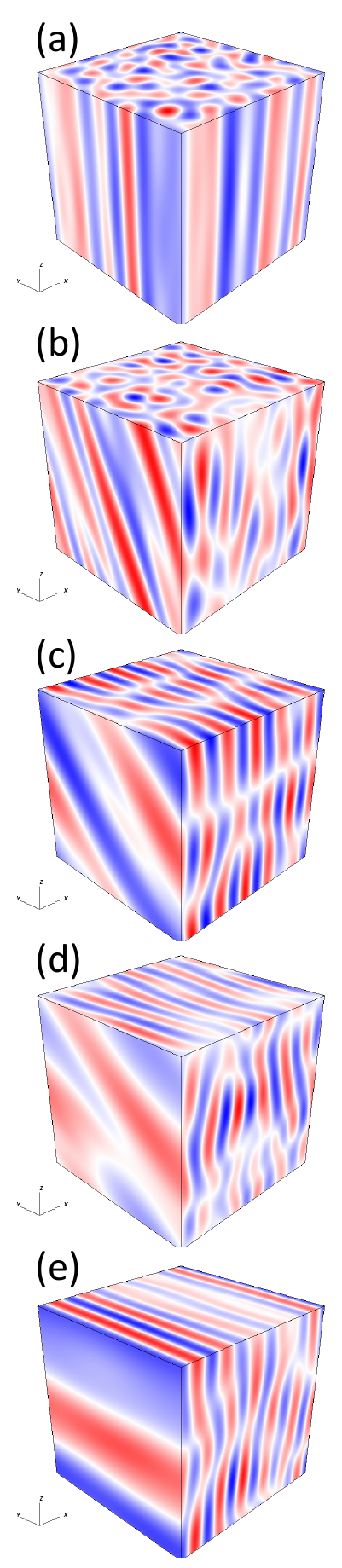}
\caption{Snapshots of the vertical velocity field during the growth of the linear instability for simulations with ${\rm Pr}=\tau=0.1$, $R_0^{-1}=1.25$, ${\rm Ta^*}=1$, and various values of $\theta$: (a) $\theta = 0$, (b) $\theta = \pi/8$, (c) $\theta = \pi/4$, (d) $\theta = 3\pi/8$ and (e) $\theta = \pi/2$. }
\label{fig:InstabSnap}
\end{figure}

Figure \ref{fig:basicInstInc} shows the growth of the linear instability by way of the heat flux as a function of time for the set of simulations with ${\rm Ta^*}=10$. Each simulation grows at roughly the same rate regardless of inclination, which is expected from linear theory. However there is a slight difference between the amplitudes in inclined simulations and the non-inclined simulation. This can be understood by considering that the simulations are initialized with small amplitude, random perturbations on the grid scale and many more modes are initially attenuated in the inclined case than in the case where $\theta=0$ (see Section \ref{sec:LinStab}). As a result the amount of energy in the initial perturbations projected onto the fastest growing modes is smaller.
Figure \ref{fig:InstabSnap} shows snapshots of the chemical composition field during the growth of the primary instability for simulations with $\theta = 0$, $\frac{\pi}{8}$, $\frac{\pi}{4}$, $\frac{3\pi}{8}$, and $\frac{\pi}{2}$. In all inclined simulations, there are prominent modes that are invariant in the direction of rotation. In simulations with smaller (or no) inclinations ($\theta = 0$ and $\frac{\pi}{8}$) the dominant modes are those with structure both in the $x$ and $y$ directions while simulations with larger inclinations (simulations that are closer to the equator) have a strong preference for modes that are invariant in the plane spanned by the rotation and gravity vectors. 

While the behavior of the linearly unstable phase is qualitatively similar for both low and high ${\rm Ta^*}$ simulations regardless of $\theta$, we find that this is not the case after the saturation of the basic instability. While inclination has only small effects on systems in the low ${\rm Ta^*}$ regime, it has a more significant influence on post-saturation dynamics in the high ${\rm Ta^*}$ regime.

\begin{figure}[h] %fig 20
\centering
%\subfloat[]{\includegraphics[scale=0.90]{fig19a.eps} \label{fig:IncFlux10}} \\
%\subfloat[]{\includegraphics[scale=0.90]{fig19b.eps} \label{fig:IncFlux1000}}
\includegraphics[width=\linewidth]{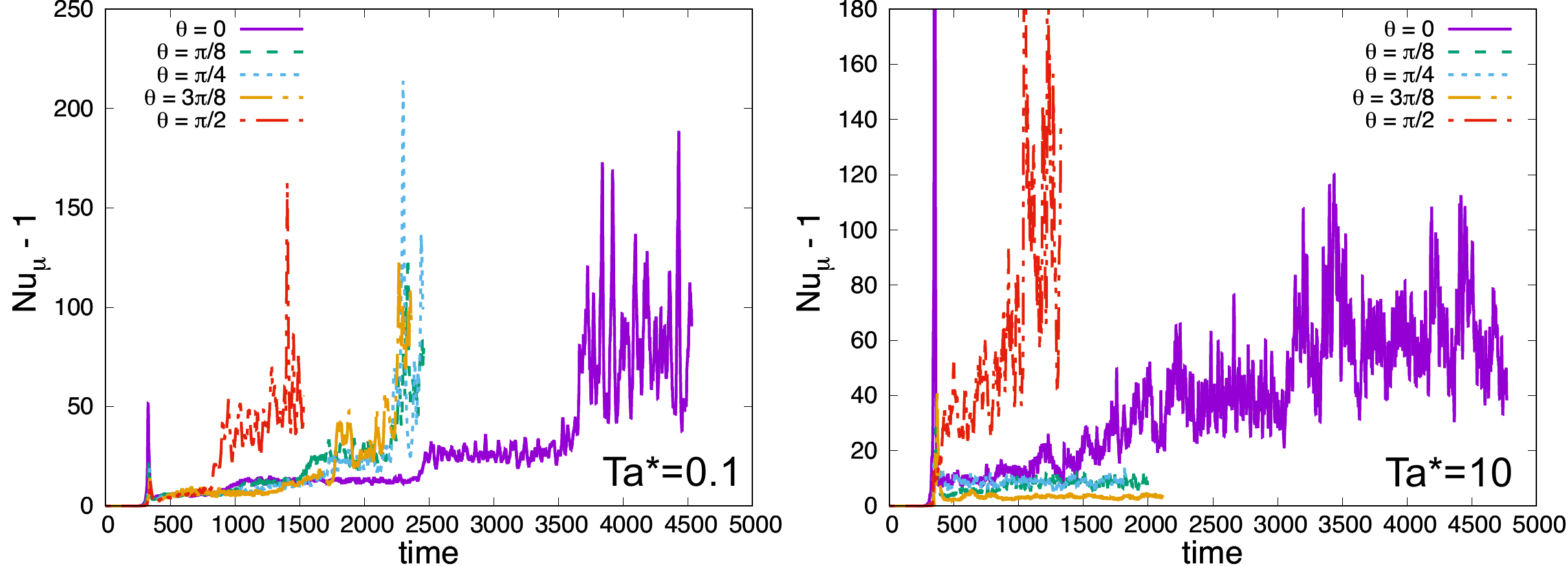}
\caption{Long-term behavior of nondimensional turbulent fluxes of composition for simulations with stated values of $\theta$ and with ${\rm Ta^*}=0.1$ (left) and ${\rm Ta^*}=10$ (right). In both sets of simulations, ${\rm Pr}=\tau=0.1$, and $R_0^{-1}=1.25$. In the low ${\rm Ta^*}$ case the succession of layered phases is similar for polar and inclined simulations, with only small differences in layering time scales and turbulent fluxes. In the high ${\rm Ta^*}$ case, fluxes in inclined simulations are sharply attenuated compared to the polar case. }
\label{fig:IncFlux}
\end{figure}

Figure \ref{fig:IncFlux} shows turbulent compositional fluxes for simulations in the low ${\rm Ta^*}$ regime (${\rm Ta^*}=0.1$). The fluxes in the homogeneously turbulent phase are roughly independent of $\theta$ indicating that inclination should have a minimal effect on the growth rate of layering modes through the $\gamma$-instability. Indeed, the stepwise increases in fluxes over time show that layer formation occurs at all inclinations. Inspection of the chemical composition profiles show that the layer interfaces are perpendicular with the direction of gravity, regardless of inclination. The latter appears to affect the layer formation timescale and layer merger rate, but this could simply be due to the inherent stochasticity of the convective layers. Finally, aside from the equatorial case, we find that inclination has a minimal impact on flux in each layered phase, so the flux laws discussed in Section \ref{sec:LowTa} apply more or less at all latitudes. As a result, we expect that heat and compositional fluxes through layered convection on a sphere should be fairly isotropic. 

Figure \ref{fig:IncFlux} also shows the turbulent compositional flux for simulations in the high ${\rm Ta^*}$ regime (${\rm Ta^*}=10$). The lack of clear stepwise increases indicates that layer formation is suppressed for most values of $\theta$ (as is the case in non-inclined simulations). The notable exception to this rule is the simulation at the equator ($\theta=\frac{\pi}{2}$) where layers are observed to form even in the high ${\rm Ta^*}$ case. Why they form in this case remains to be determined.

\begin{figure} %fig 21
\centering
%\subfloat[]{\includegraphics[scale=0.35]{vortxy1000_0125.jpg} \label{fig:IncVort_a}} \hspace{1em}
%\subfloat[]{\includegraphics[scale=0.35]{vortxy1000_025.jpg} \label{fig:IncVort_b}} \\
%\subfloat[]{\includegraphics[scale=0.35]{vortxy1000_0375.jpg} \label{fig:IncVort_c}}
\includegraphics[width=\linewidth]{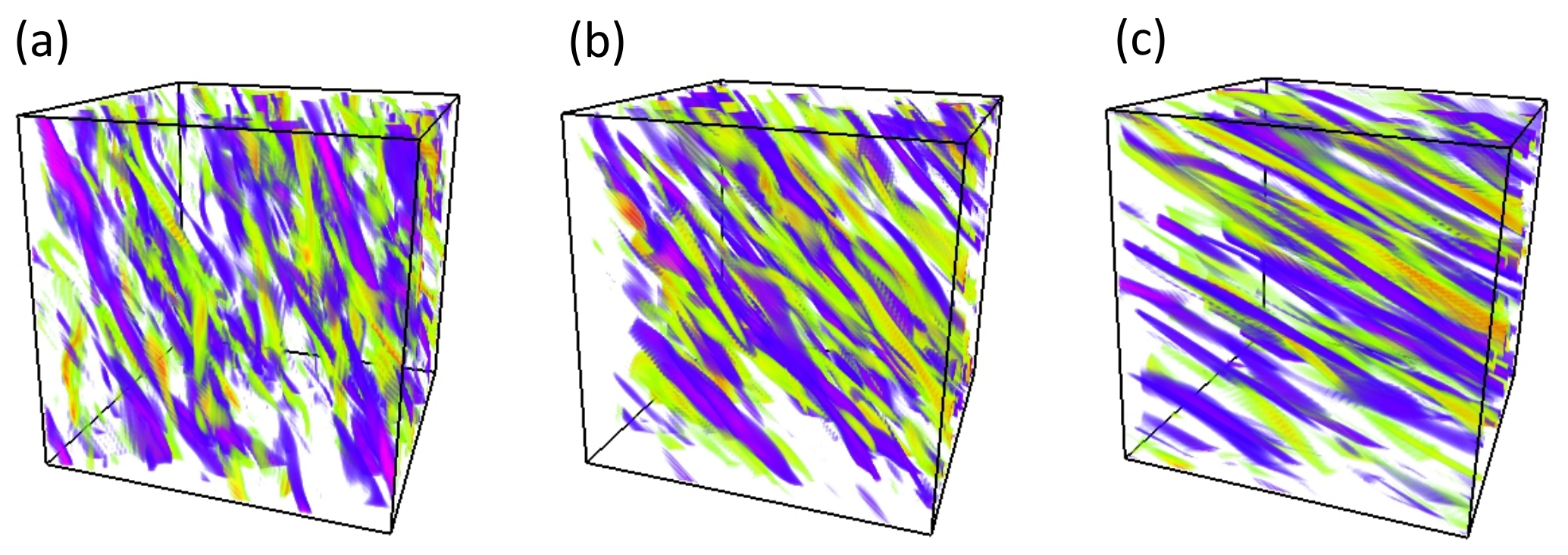}
\caption{Snapshots of $\omega_{yz}$, the component of the vorticity parallel to the rotation axis, during saturation of the linear instability. Shown are simulations with ${\rm Pr}=\tau=0.1$, $R_0^{-1}=1.25$, ${\rm Ta^*}=1$, and (a) $\theta=\frac{\pi}{8}$, (b) $\theta=\frac{\pi}{4}$ and (c) $\theta=\frac{3\pi}{8}$. In each case, coherent small scale vortices are aligned with the axis of rotation. \label{fig:IncVort}}
\end{figure}

Another major difference between inclined and non-inclined simulations is that there is no evidence for the large scale vortices in simulations with $\theta \ne 0$, even though they are observed in $\theta=0$ simulations at the same parameters (see Section \ref{sec:thetaZero}). This is illustated in Figure \ref{fig:IncVort} which shows the quantity $\omega_{yz}=\frac{ \mathbf{\omega} \cdot \mathbf{\Omega} }{\left| \mathbf{\Omega} \right|}$. There are many smaller scale vortices aligned with the rotation axis but no large scale vortex. This is even true in the simulation with the smallest inclination ($\theta=\frac{\pi}{8}$), bringing into question whether large scale vortices would be common in stars and planets except exactly at the poles. The inclination of the small scale vortices is associated with smaller vertical transport, and Figure \ref{fig:IncFlux} suggests that mixing becomes less efficient as $\theta$ gets larger (except very close to the equator). %The dichotomy between slowly rotating simulations where inclination is not an important effect suggests that there is the potential for observable 

%CONCLUSION
\section{Conclusion} \label{sec:conclusion}
\subsection{Summary and discussion}
The main result of this study is the discovery of two distinct regimes in rotating ODDC depending on whether the rotation rate $\Omega$ is high or low. We find that the most appropriate parameter for determining if a system is in one regime or the other is ${\rm Ta^*} =\frac{4 \Omega^2 d^4}{\kappa_T^2}$, where $d$ is given in Equation (\ref{eq:nondim}) and $\kappa_T$ is the thermal diffusivity. The transition from the regime with slow rotation to the regime that is rotationally dominated occurs consistently at ${\rm Ta^*} \approx 1$. 

In the low $\rm{Ta^*}$ regime in polar regions (with $\theta=0$), rotating ODDC behaves in a qualitatively similar way to non-rotating ODDC. The transition to layered convection (or lack thereof) at low $\rm{Ta^*}$ is consistent with the predictions of $\gamma$-instability theory made for non-rotating ODDC \citep{Mirouh2012}: at parameters where layers form in non-rotating ODDC, we also observe layer formation in low $\rm{Ta^*}$ simulations. Likewise, in the simulations we ran at non-layered parameters, we find that low $\rm{Ta^*}$ simulations do not form layers, and are dominated by gravity waves like their non-rotating counterparts \citep{Moll2016}. We understand this to be true because the thermal and compositional fluxes immediately after saturation of the primary instability of ODDC are unaffected by rotation in this regime. Since the $\gamma$-instability only depends on these fluxes, it is similarly unaffected. Given the limited number of available simulations, we cannot say definitively what effect (if any) rotation has on the layering threshold $R_L^{-1}$ (the value of the inverse density ratio, $R_0^{-1}$, below which layers are predicted to form through the $\gamma$-instability, and above which they are not), only that it is not significant in the low $\rm{Ta^*}$ simulations presented here, which are far from that threshold. However, we believe that $R_L^{-1}$ would be relatively unaffected by rotation for ${\rm Ta^*} < 1$.

Beyond these qualitative similarities with non-rotating simulations, however, rotation in low $\rm{Ta^*}$ simulations has a deleterious effect on thermal and compositional transport in both the layered and non-layered parameter regimes. For a given layer height, turbulent fluxes through a thermo-compositional staircase decrease as rotation increases (eg. by about $50\%$ in the $\rm{Ta^*}=0.1$ simulation presented in Section \ref{sec:LowTa}). However, our results also suggest that this effect becomes smaller as the layer height increases (through mergers, for example). For reasonably large layer heights, we postulate that rotation has a minimal effect on ODDC, and that the flux laws originally proposed by \citet{Wood2013} actually hold. Turbulent fluxes through non-layered ODDC in the gravity-wave-dominated phase are reduced by as much as $90\%$ compared with the non-rotating case, but this merely implies that they remain negligible as discussed by \citet{Moll2016}.

Finally, low $\rm{Ta^*}$ simulations at higher colatitude $\theta$ are not significantly different from their polar counterparts. Inclination has a negligible effect on the temperature and compositional fluxes, but may induce differences in the time scales of layer formation and mergers.

In the high ${\rm Ta^*}$ regime, dynamics are radically different from non-rotating and low ${\rm Ta^*}$ simulations. Most striking is that layer formation is inhibited at low inverse density ratios (except in the equatorial case). Instead, the dynamics are dominated by vortices aligned with the direction of rotation, which span the domain. Their horizontal scales seem to depend on $R_0^{-1}$, $\theta$ and ${\rm Ta^*}$. In polar regions, we observe that some simulations become dominated by a single large scale cyclonic vortex which grows to fill the domain, similar to those observed by \citet{guervilly2014} in rotating Rayleigh-B\'enard convection. Our preliminary data show that this phenomenon may be limited to low $R_0^{-1}$, together with ${\rm Ta^*}$ between $1$ and $10$, but the precise conditions necessary for these large scale vortices to form remain to be determined. We find that large scale vortices do not occur in the most rapidly rotating simulation (${\rm Ta^*} = 90$) at low $R_0^{-1}$, in any of the high $R_0^{-1}$ simulations or in any of the inclined simulations. In these cases the system dynamics are instead dominated by many smaller scale vortices of both polarities.

Turbulent fluxes through different types of vortices vary with parameters in a complex manner, and it is therefore difficult to make general statements about them. The fluxes in the presence of large scale vortices are significant, but appear to be highly dependent on the dimensions of the domain, which makes it difficult to predict in situ mixing in a star or planet. On the other hand, fluxes in the presence of small scale vortices are not likely to be dependent on domain size, but their dependence on ${\rm Ta^*}$ and $R_0^{-1}$ has yet to be extensively studied. The most definitive aspect of the fluxes in simulations that host small scale vortices is that they are highly dependent on the inclination $\theta$, with higher inclinations causing less efficient transport (except at the equator). This is due to the fact that velocities are constrained to being along the axis of rotation by Taylor-Proudman effects. It is interesting to note that in the high ${\rm Ta^*}$ regime, layers form in our equatorial simulation ($\theta=\frac{\pi}{2}$). In this run, turbulent fluxes through the layers are comparable to layered fluxes in the low ${\rm Ta^*}$ and non-rotating regimes. All of this suggests that the poles and equator may be regions of strongly enhanced temperature and compositional transport in ODDC, while turbulent mixing at latitudes in between is quenched.

Finally, simulations where ${\rm Ta^*} \approx 1$ appear to be edge cases with features of both high and low ${\rm Ta^*}$. At parameters conducive to layering, simulations with ${\rm Ta^*} \approx 1$ show evidence of perturbations to the background density profiles, indicating the growth of the $\gamma$-instability. However, we also see the development of large-scale vertically invariant vortices which prevent actual layered convection from occurring. Also, when ${\rm Ta^*} \approx 1$ at non-layered, gravity-wave-dominated parameters,  we see evidence of gravity waves as well as small thin vortices which are nearly vertically invariant.

%CAVEATS
There are several caveats to these conclusions that should be mentioned. The dimensionality of parameter space that would need to be explored to provide a comprehensive study of rotating ODDC is high, and comprises of $(L_x,L_y,L_z)$, $\theta$, ${\rm Pr}$, $\rm \tau$, ${\rm Ta}^*$ and finally $R_0^{-1}$. Computational limitations have forced us to be highly selective on the sets of simulations explored so this study does not constitute a comprehensive sweep of parameter space. As such, there may be behaviors that occur at unexplored parameters that are not addressed here. 

First of all, in the interest of reducing computational expense we have chosen to run most of our simulations in domains with dimensions $(100d)^3$. With boxes of this size layers always merge until a single interface remains (as in the non-rotating case). It would be interesting to see in a taller domain if rotation has a role in determining the layer height (ie. to see if layers stop merging before reaching the one-layered phase). Wider domain sizes may also help to answer questions about the vortices present in high ${\rm Ta^*}$ simulations. Particularly, they may reveal whether large scale vortices have a characteristic horizontal length, or whether they always grow to fill the domain. Wider boxes may also show if there are so far undetected large scale features emerging in systems dominated by small scale vortices.

Another area of uncertainty is that the chosen values of $R_0^{-1}$, $R_0^{-1} = 1.25$ and $R_0^{-1} = 4.25$, are fairly close to the convective and marginal stability thresholds, respectively, making them somewhat extreme cases. While we do not believe that choosing less extreme parameter values would lead to dramatic qualitative changes in the results, we cannot rule this possibility out until further work has been completed.

Finally, for computational reasons, the values of ${\rm Pr}$ and $\tau$ chosen for our simulations (${\rm Pr}=\tau=0.1$ and $0.3$) are substantially larger than the values in stellar interiors (where ${\rm Pr}\sim\tau\sim10^{-6}$) and the interiors of giant planets like Jupiter and Saturn (where ${\rm Pr}\sim\tau\sim10^{-3}$). Consequently, there may be additional physical effects that occur at low parameter values that are not observed here. However, the values used here may be closer to actual values for ice giants such as Uranus and Neptune whose equations of state are influenced by the presence of water and methane ices in their atmospheres \citep{Redmer2011}.

%Other caveats to the analysis presented here is that the Boussinesq approximation was used and that other physical effects such as large scale shear and magnetic fields were ignored.

%Prospects for stellar and planetary modeling
\subsection{Prospects for stellar and planetary modeling}
As summarized above, our results for the low ${\rm Ta^*}$ regime show that attenuation of the fluxes due to rotational effects is not likely to be significant for astrophysical models or observations. In this regime we advocate use of the parameterizations presented in \citet{Wood2013} in layered ODDC and \citet{Moll2016} in non-layered ODDC. However, a potentially observable effect of rotating ODDC in this regime could be related to how rotation affects the structure of layers and interfaces in low ${\rm Ta^*}$ simulations. We have found in our experiments that rotation leads to layer interfaces that are more stably stratified than in non-rotating simulations. It may be possible in the future that such steep density gradients could be observed in a star through asteroseismology. This line of inquiry could also be extended to gas giant planets. In Saturn, for example, it may be possible to detect density gradients using ring seismology \citep{Fuller2014}, and in the future, we may even be able to probe the interior structure of Jupiter through detection of global modes \citep{gaulme2011}.

In the high ${\rm Ta^*}$ regime the results of this study have potential observational implications for thermal and compositional transport in stars and planets. The sensitivity of the turbulent heat flux to the inclination in our high ${\rm Ta^*}$ simulations suggests that the transport in a rapidly rotating giant planet could vary substantially with latitude (with higher fluxes at the poles and equator). Indeed, the gas giants in our own solar system are found to have luminosities that are independent of latitude, despite the fact that regions close to the equator receive more solar energy. Since we would expect regions of Jupiter or Saturn's atmosphere that get more radiation from the sun to have higher luminosities (because they are reradiating more solar energy) the isotropy of the outgoing flux in luminosity suggests that more heat from the interiors of these planets is being radiated at the poles than at other latitudes. Further study is warranted to determine if rapidly rotating ODDC contributes to this effect. The large-scale vortices present in the polar regions of the high ${\rm Ta^*}$ simulations present an intriguing observational potential, of regions with strong heat and compositional fluxes and strong collimated vertical flows. However, there is reasonable doubt as to whether these large scale vortices represent a real physical phenomenon. We only observe them to occur in polar simulations (in a limited range of ${\rm Ta^*}$), and it is possible that even a slight misalignment between the direction of gravity and the rotating axis could prevent them from forming.

For stars, in the case of semi-convection zones adjacent to convection zones, \citet{Moore2015} showed that non-rotating ODDC is always in the layered regime, and that transport through the semi-convective region is so efficient that the latter gets rapidly absorbed into the convection zone. In essence, aside from a fairly short transient period, the star evolves in a similar way taking into account semi-convection, or ignoring it altogether and using the Schwarzchild criterion to determine the convective boundary. Our results suggest that this conclusion remains true for slowly rotating stars. However, if the star is in the high ${\rm Ta^*}$ regime instead, layered convection is suppressed, transport through the semi-convective region is much weaker, and may possibly depend on latitude. This would in turn imply fairly different evolutionary tracks and asteroseismic predictions.

\acknowledgements
This work was funded by NSF-AST 1211394 and NSF-AST 1412951. The authors are indebted to Stephan Stellmach for granting them the use of his code, and for helping them implement the effects of rotation. 

\appendix
\section*{Appendix: Minimum mode amplitudes for layered convection} \label{sec:appA}
In rotating systems density perturbations must grow to a higher amplitude in order for layered convection to occur, compared to non-rotating ones. This can be understood better by considering how convective plumes form at the edges of the diffusive boundaries in layered convection. In order for a hot plume at the bottom of a layer to rise it must displace the fluid above it. Because the interfaces act as flexible but more-or-less impenetrable boundaries, fluid that is moving upward because of the rising plume must be deflected by the top boundary and displaced horizontally. Rotation resists motion involving gradients of velocity in the direction of the rotation axis. In order to overcome this resistance, and therefore for convection to take place in the layer, a more strongly positive density gradient must be present between the interfaces.

We make quantitative estimates of this effect on layered convection in ODDC through adaptation of a theory related to rotating Rayleigh-B\'enard convection \citep{chandrasekhar1961}. Assuming free boundary conditions, the critical Rayleigh number, ${\rm Ra}_c$, for rotating Rayleigh-B\'enard convection is the following function of ${\rm Ta^*}$:
\begin{equation} \label{eq:RaCritAp}
{\rm Ra}_c({\rm Ta}^*) = 3\pi^4 \left( \frac{H^4 {\rm Ta}^*}{2{\rm Pr}^2\pi^4} \right)^{\frac{2}{3}} + \frac{27\pi^4}{4} \, ,
\end{equation}
where $H$ is the layer height. In our non-dimensionalization the critical density gradient for a convective layer, $\left| \frac{\partial \rho}{\partial z} \right|_c$, can be written in terms of ${\rm Ra}_c$ as
\begin{equation} \label{eq:RaDimConstAp}
\left| \frac{\partial \rho}{\partial z} \right|_c = \frac{ {\rm Ra}_c }{H^4} \, .
\end{equation}
Then by considering a density profile defined as
\begin{equation} \label{eq:DensProfAp}
\rho  = (1-R_0^{-1})z + 2 A_n \sin{\left( k_n z \right)} \, ,
\end{equation}
where $A_n$ is the amplitude of the perturbation from a single layering mode with vertical wavenumber $k_n$, we provide a second definition for the critical density gradient
\begin{equation} \label{eq:DensGradAp}
\left| \frac{\partial \rho}{\partial z} \right|_c = \max{\left( \frac{\partial \rho}{\partial z} \right)} = 1-R_0^{-1} + 2 A_n k_n \, .
\end{equation}

From equations (\ref{eq:RaCritAp}) through (\ref{eq:DensGradAp}) we can then generate an expression for $\left| A_n \right|$ in terms of ${\rm Ta}^*$, $R_0^{-1}$, $H$, and $k_n$, and thus get an estimate for the critical layering mode amplitude for the onset of layered convection,
\begin{equation} \label{eq:ModeAmpAp}
\left| A_n \right| = \left| \frac{\frac{{\rm Ra}_c}{H^4} + \left(R_0^{-1} - 1\right)}{2 k_n} \right| = \left| \frac{ \frac{3\pi^4}{H^4}\left(\frac{H^4{\rm Ta}^*}{2{\rm Pr}^2\pi^4}\right)^{\frac{2}{3}} + \frac{27\pi^4}{4H^4} + \left( R_0^{-1} - 1 \right) }{2k_n} \right| \, .
\end{equation}
This formula recovers Equation (29) of \citet{rosenblum2011} in the non-rotating limit, as long as the term $27 \pi^4/4H^4$ can be neglected (which is always true for physically realizable layer heights, that typically have $H > 30$).

\bibliographystyle{apj}
%\bibliography{DDC_bib.bib}

\end{document}